\documentclass[runningheads]{llncs}

\usepackage[T1]{fontenc}
\usepackage[utf8]{inputenc}

\usepackage{cite}
\usepackage{amsmath,amssymb,amsfonts}
\usepackage{algorithm}
\usepackage{algorithmic}
\usepackage{graphicx}
\usepackage{xcolor}
\usepackage{array}
\usepackage{booktabs}
\usepackage{tabularx}
\usepackage{threeparttable}
\usepackage{enumitem}
\usepackage{verbatim}
\usepackage{url}
\usepackage[hidelinks]{hyperref}
\usepackage[capitalize]{cleveref}
\usepackage{listings}
\usepackage[most]{tcolorbox}
\usepackage{bbding}
\usepackage{microtype}

\crefname{figure}{Fig.}{Figs.}
\Crefname{figure}{Fig.}{Figs.}
\crefname{section}{Section}{Sections}
\Crefname{section}{Section}{Sections}

\crefname{listing}{Listing}{Listings}
\Crefname{listing}{Listing}{Listings}
\crefname{lstlisting}{Listing}{Listings}
\Crefname{lstlisting}{Listing}{Listings}

\lstdefinestyle{mystyle}{
    basicstyle=\ttfamily\footnotesize\linespread{0.88}\selectfont,
    breakatwhitespace=false,
    breaklines=true,
    captionpos=b,
    keepspaces=true,
    numbers=left,
    numbersep=10pt,
    showspaces=false,
    showstringspaces=false,
    showtabs=false,
    tabsize=2,
    framexleftmargin=15pt,
    xleftmargin=15pt,
    morekeywords={f0, f1, f2, topic0, topic1, topic2, MulRecv, MulSend, VecToPubs, VecToSubs},
    keywordstyle=\color{red}\bfseries,
}

\newtcolorbox{codebox}[1][]{
    enhanced,
    unbreakable,
    colback=white,
    colframe=black,
    boxrule=0.5pt,
    sharp corners,
    boxsep=2pt,
    left=5pt,
    right=5pt,
    top=2pt,
    bottom=2pt,
    listing only,
    listing options={style=mystyle},
    overlay={\node[anchor=south east, outer sep=2pt, font=\footnotesize] at ([xshift=-3pt, yshift=3pt]frame.south east) {#1};}
}

\newif\ifanonymous
\anonymousfalse

\makeatletter
\renewenvironment{table}
{\setlength\abovecaptionskip{0\p@}%
    \setlength\belowcaptionskip{2\p@}%
    \@float{table}}
{\end@float}
\renewcommand\section{\@startsection{section}{1}{\z@}%
    {-12\p@ \@plus -4\p@ \@minus -4\p@}%
    {8\p@ \@plus 4\p@ \@minus 4\p@}%
    {\normalfont\large\bfseries\boldmath
        \rightskip=\z@ \@plus 8em\pretolerance=10000 }}
\renewcommand\subsection{\@startsection{subsection}{2}{\z@}%
    {-12\p@ \@plus -4\p@ \@minus -4\p@}%
    {5\p@ \@plus 4\p@ \@minus 4\p@}%
    {\normalfont\normalsize\bfseries\boldmath
        \rightskip=\z@ \@plus 8em\pretolerance=10000 }}
\renewcommand\subsubsection{\@startsection{subsubsection}{3}{\z@}%
    {-12\p@ \@plus -4\p@ \@minus -4\p@}%
    {-0.5em \@plus -0.22em \@minus -0.1em}%
    {\normalfont\normalsize\bfseries\boldmath}}
\makeatother
\setlist{topsep=2pt,itemsep=0pt,parsep=0pt,partopsep=0pt}
\graphicspath{{figures/}}
\begin{document}

\title{Awkernel: RTOS Bridging DAG Scheduling Theory and Component-Oriented Real-Time Systems Practice}
\titlerunning{Awkernel: Bridging DAG Scheduling Theory and Practice}

\ifanonymous
    \author{Anonymous Author(s)}
    \authorrunning{Anonymous Author(s)}
    \institute{Anonymous Institute}
\else
    \author{Atsushi Yano\inst{1,2}\textsuperscript{(\Envelope)}\orcidID{0000-0003-2420-3622} \and
        Ryo Nishimura\inst{1} \and
        Yutaro Kobayashi\inst{2} \and
        Yuuki Takano\inst{2} \and
        Takahiro Ishikawa-Aso\inst{3,2} \and
        Takuya~Azumi\inst{4}\orcidID{0000-0003-0767-4086}}
    \authorrunning{A. Yano et al.}

    \institute{Graduate School of Science and Engineering, Saitama University, Saitama, Japan\\
        \email{yano.a.112@ms.saitama-u.ac.jp}
        \and
        TIER IV, Inc., Tokyo, Japan
        \and
        The University of Tokyo, Tokyo, Japan
        \and
        Academic Association (Graduate School of Science and Engineering), Saitama University, Saitama, Japan}
\fi

\maketitle

\begin{abstract}

    Real-time directed acyclic graph (DAG) scheduling provides a rigorous basis for timing analysis and efficiency in component-oriented real-time systems (CORTS) such as autonomous driving systems, yet challenges remain from the perspectives of both theorists and practitioners.
    For theorists, no OS natively supports the DAG task model; algorithms are therefore evaluated only analytically or in simulation, ignoring real-hardware overheads.
    For practitioners, today's CORTS platforms (ROS~2 and AUTOSAR Adaptive Platform) are designed for service composition rather than for timing analyzability; adopting DAG scheduling research therefore demands substantial code changes and developer discipline.
    This paper proposes Awkernel, the first open-source real-time operating system (RTOS) that natively supports real-time DAG scheduling research, bridging DAG scheduling theory and CORTS practice.
    For theorists, Awkernel is a real-hardware testbed: it exposes a modular OS-native DAG scheduler for easy algorithm implementation, plus a random workload generator and built-in metrics for evaluation.
    For practitioners, Awkernel is not a full replacement for today's platforms but a CORTS platform maximizing timing analyzability: its familiar publish/subscribe model keeps porting mechanical, mapping each callback to one subtask.
    Moreover, its function-as-subtask (FasS) API enforces DAG semantics (precedence constraints) without programmer discipline, yielding applications that match the theoretical DAG model and directly inherit its efficiency and safety guarantees.
    On a Raspberry Pi~4, microbenchmarks confirm microsecond-level scheduling overheads; a global earliest deadline first (GEDF) scheduler fits in 226~lines; and a practitioner ports part of an autonomous driving system using only the FasS API, without reasoning about DAG semantics.

    \keywords{Operating Systems \and Real-time Scheduling \and Publish/Subscribe \and Directed Acyclic Graph}

\end{abstract}

\section{Introduction}\label{sec:intro}

Modern cyber-physical systems, exemplified by autonomous driving systems, are typically realized as \emph{component-oriented real-time systems} (CORTS) in which a collection of units implementing minimal functionalities cooperatively achieve the system-level objectives through mutual message-passing communication~\cite{FasS, CIE}.
Representative CORTS platforms include ROS~2~\cite{ros2_repo, ros2_paper} (used in autonomous driving stacks such as Autoware~\cite{autoware}) and the AUTOSAR Adaptive Platform (AP)~\cite{fuerst-autosar-ap-2016} (the automotive-industry standard for high-performance electronic control units (ECUs)).
Both are POSIX-based middleware layers that delegate scheduling decisions to the underlying OS~\cite{henle-ros2-vs-ap-2022, bellassai-slelet-2025}.
Each unit is implemented as callbacks triggered by timers or incoming messages that publish to downstream channels, forming a system-wide dataflow whose timing requirements the underlying platform must satisfy efficiently.

Real-time directed acyclic graph (DAG) scheduling offers a rigorous foundation for timing analysis in such systems by explicitly representing intra-task parallelism and dependencies~\cite{verucchi2023survey}.
Yet no publicly available OS or runtime currently treats the DAG task model as a first-class OS abstraction, that is, a concept the kernel itself represents in its internal data structures, exposes through dedicated scheduling APIs, and whose semantic constraints are enforced by the OS scheduler rather than by a user-space runtime.
DAG-aware techniques (global earliest deadline first (GEDF)~\cite{GEDF}, reservation-based scheduling~\cite{holistic}, fluid scheduling~\cite{guan-2022}, and adaptive scheduling~\cite{adaptive1}) therefore cannot run on real hardware, and proposed DAG algorithms are evaluated only by analysis or simulation, hiding overheads that only real hardware exposes.

Although a ROS~2 system superficially resembles a DAG, its execution semantics diverge from the theoretical DAG model due to middleware-level behaviors such as executor-specific callback scheduling~\cite{casiniros, ros_stv, ros_multi} and customizable message synchronization~\cite{ros_sync}.
A large body of work has analyzed and improved the timing behavior of ROS~2 executors~\cite{casiniros, ros_stv, ros_multi, ros_e2e, ros_sync, ros_single_pri}, but the resulting two-level scheduling complicates analysis and yields ROS~2-specific models that cannot reuse the DAG scheduling literature at the OS scheduling layer.
This is also a practical concern: modern CORTS such as autonomous driving stacks already carry substantial application-level complexity, so it is preferable that the platform beneath them contribute as little complexity as possible to the timing argument on which safety assurance rests. From this perspective, scheduling behavior arising from the interaction of two independently designed layers leaves room for improvement.
Efforts to apply DAG scheduling directly to ROS~2~\cite{CIE, multi_dead, FasS} face two further obstacles: callback description is free enough to break strict DAG semantics~\cite{FasS}, and no platform offers runtime control based on DAG information.
Consequently, CORTS practitioners cannot readily exploit the rich legacy of real-time DAG scheduling.

\textbf{Research questions (RQ):} Building on the above background, this work formulates the following three research questions:

\vspace{-0mm}

\begin{itemize}
    \item[1.] What differences exist between the DAG model and the models adopted by representative CORTS platforms, and what requirements must system software satisfy to bridge these differences and enable DAG scheduling on real hardware?
    \item[2.] How can DAG scheduling theorists evaluate their algorithms on real hardware with low implementation effort?
    \item[3.] How can practitioners of CORTS leverage the existing body of research on DAG scheduling?
\end{itemize}

\vspace{-0mm}

\textbf{Awkernel} is the first open-source real-time operating system (RTOS) that natively supports real-time DAG scheduling research, bridging DAG scheduling theorists and practitioners of CORTS.
For theorists, it is a real-hardware testbed for DAG scheduling algorithms; for practitioners, it is not a complete replacement for ROS~2 but a CORTS platform maximizing timing analyzability and direct reuse of the DAG scheduling literature.
Concretely, Awkernel (i) is open-source, unlike commercial AP implementations; (ii) treats DAG structure as a first-class OS abstraction, unlike platforms that delegate scheduling to a generic POSIX OS; and (iii) exposes a scheduler interface for implementing, swapping, and measuring DAG scheduling algorithms on real hardware.
Awkernel provides a ROS~2-like publish/subscribe model and the function-as-subtask (FasS) API~\cite{FasS}, an OS-level interface for native DAG scheduling that represents each subtask as a function whose arguments and return values are the input and output edges of the DAG. By minimizing the freedom of subtask description, the OS enforces DAG semantics without relying on programmer discipline.
Awkernel further pairs a scheduler with arbitrary preemption granularity, a DAG management substrate tracking runtime metadata, and a scheduler interface decoupling algorithms from OS internals; practitioners apply any DAG scheduling algorithm via the FasS API, and theorists evaluate algorithms on real hardware with minimal effort.
It further includes a random DAG workload generator built on RD-Gen~\cite{rd-gen} and in-kernel metrics covering context-switch counts, preemption overhead, and DAG response times, enabling comprehensive evaluation of algorithm generality and performance.

\textbf{Contributions:} The main contributions of this paper include the following:
\vspace{-0mm}
\begin{itemize}
    \item We analyze the structural differences between the theoretical DAG model and POSIX-based CORTS platforms (ROS~2 and AUTOSAR Adaptive Platform), derive the conditions and abstractions required to realize DAG scheduling on real hardware, and confirm Awkernel satisfies them.
    \item The integration of a scheduler that supports arbitrary preemption granularity, together with a DAG management substrate, the FasS API, and a random DAG workload generation facility into Awkernel establishes a comprehensive platform. This environment facilitates the straightforward implementation of DAG scheduling algorithms and enables evaluation on real hardware.
    \item The combination of the publish/subscribe communication model within Awkernel and the FasS API enables practitioners of CORTS to leverage real-time DAG scheduling with modest, largely mechanical porting effort.
\end{itemize}
\vspace{-0mm}

\begin{table}[t]
    \caption{Evaluation methods of DAG scheduling papers from the top real-time systems conferences over the past 15 years.}
    \vspace{-0mm}
    \label{tab:dag_evaluation}
    \centering
    \scriptsize
    \begin{tabularx}{\linewidth}{|X|c|c|c|}
        \hline
        \textbf{References}                                                                                                       & \textbf{Simulation} & \textbf{Schedulability Tests} & \textbf{Real Hardware} \\
        \hline
        \cite{han,pass,guo,petros,niklas-2023,yue-2023,exact,qingqiang,multi_dead,kunal-2018}                                     & \checkmark          &                               &                        \\
        \hline
        \cite{junjie,saud,ahmed,guan,ching-chi,zheng,casini-2018,xu-2017,xu-2016,li-2014,qingqiang-2021,pathan-2018,aromolo-2023} &                     & \checkmark                    &                        \\
        \hline
        \cite{micaela,holistic,xu-2021,shuai}                                                                                     & \checkmark          & \checkmark                    &                        \\
        \hline
    \end{tabularx}
    \vspace{-0mm}
\end{table}

\section{Related Work}\label{sec:related}
\vspace{-0mm}
\subsection{Theoretical Real-Time DAG Scheduling}
\vspace{-0mm}
Theoretical real-time DAG scheduling addresses computational complexity, resource heterogeneity, and multicore timing guarantees across autonomous, automotive/avionics, edge/IoT, and high-performance computing workloads, yet how these advances are validated in practice remains unclear.
To address this question, we survey 27 papers on real-time scheduling for DAG tasks from ECRTS, RTSS, and RTAS over the past 15 years~\cite{han,pass,guo,petros,niklas-2023,yue-2023,exact,qingqiang,multi_dead,kunal-2018,junjie,saud,ahmed,guan,ching-chi,zheng,casini-2018,xu-2017,xu-2016,li-2014,qingqiang-2021,pathan-2018,aromolo-2023,micaela,holistic,xu-2021,shuai}.
\cref{tab:dag_evaluation} shows all rely on simulations and schedulability tests; none use real hardware. This reinforces the problem in~\cref{sec:intro}.

The only widely accepted real-time scheduling testbed, LITMUS${ }^{\text{RT}}$~\cite{litmus-rt}, is a Linux-based plugin framework for empirically comparing partitioned, global, and Pfair multiprocessor algorithms under realistic overheads (scheduling latencies, context-switch costs), but it targets independent sequential tasks and lacks DAG support; no DAG-scheduling testbed exists, a gap Awkernel fills. Awkernel continues this testbed lineage, extending the target model to DAG tasks.

\subsection{Real-Time Support in CORTS Platforms}\label{subsec:rw_cort}
A \emph{CORTS platform} is a service-oriented middleware atop a POSIX OS, where modular components communicate by asynchronous message passing. Its two dominant instances, ROS~2 and AP, share a Data Distribution Service (DDS)/SOME/IP-on-POSIX architecture~\cite{henle-ros2-vs-ap-2022, fuerst-autosar-ap-2016}.

The single-threaded executor of ROS~2 non-preemptively dispatches callbacks in fixed-priority order~\cite{casiniros}, diverging from classical real-time schedulers and prompting dedicated analyses: response-time analysis for single-threaded~\cite{casiniros,ros_single_pri,ros_stv} and multi-threaded~\cite{ros_multi,RE14} executors, end-to-end metrics such as reaction time and data age~\cite{ros_e2e}, and DDS-layer synchronization bounds~\cite{ros_sync}.
Motivated by these analyses, numerous enhancements have been proposed to improve the real-time performance of ROS~2: chain-aware fixed-priority scheduling (PiCAS~\cite{picas}), dynamic-priority earliest deadline first (EDF) executors~\cite{RE10}, lock-free multi-threaded execution (RTeX~\cite{songran_liu_641693d9}), reconciliation with classical non-preemptive scheduling~\cite{harun_teper_44b003f3}, flexible scheduling policies (ROS\textsuperscript{RT}~\cite{ros-rt}), fully preemptive per-callback thread designs~\cite{physics_informed_mixed_criticality_scheduling_for_f1tenth_cars_with_preemptable_ros_2_executors_31f450b3,CIE}, and zero-copy inter-process communication (IPC) via shared memory (Agnocast~\cite{agnocast}).

The AUTOSAR Adaptive Platform, introduced in 2017 for advanced driver-assistance and autonomous driving ECUs~\cite{fuerst-autosar-ap-2016}, layers SOME/IP and ara::com over POSIX (Linux/QNX). Lacking timing determinism, it has spawned middleware research: SL-LET~\cite{bellassai-slelet-2025}, Logical Execution Time on POSIX~\cite{bellassai-let-posix-2023}, and response-time analysis of QNX Adaptive Partitioning~\cite{dasari-qnx-2022}.

YASMIN~\cite{yasmin} is an open-source real-time middleware natively supporting a DAG task model with FIFO-channel precedence, and global and partitioned variants of EDF and fixed-priority scheduling. However, as a user-space C library on POSIX~\cite[\S3]{yasmin}, thread-scheduling ultimately falls to the POSIX scheduler, which lacks any notion of DAG precedence, the same two-level structure as ROS~2 and AP; \cref{sub:ros2_limits} quantifies the cost of this structure empirically.

None of ROS~2, AP, or YASMIN treats the DAG model as a first-class OS abstraction, leaving structural gaps that block practitioners from leveraging real-time DAG scheduling research (see~\cref{sec:gap}).
\subsection{Real-Time Parallel Programming Models}\label{subsec:rw_openmp}
OpenMP is the de facto programming model for precedence-constrained parallel tasks within a single process; real-time research on it includes task-graph derivation (ompTG~\cite{sun-omptg-2022}), worst-case response-time analyses for OpenMP with loops~\cite{sun-openmp-wcrt-2021} and conditional branches~\cite{sun-openmp-conditional-2020}, and predictable parallel programming methodologies~\cite{serrano-openmp-2022}.
These efforts share the goal of Awkernel but operate at a different granularity: OpenMP research addresses intra-process parallelism (subtasks sharing one address space under a runtime library), whereas Awkernel targets CORTS, where independent components cooperate via publish/subscribe (e.g., ROS~2-based autonomous driving systems).

\section{Differences between DAG Model and ROS~2 Model}\label{sec:gap}
This paper focuses on ROS~2 within existing CORTS platforms because its implementation is open for API-level analysis and it dominates timing analyses at top real-time venues. AP is restricted by licensing and tooling~\cite{iwakami-ap-ros2-2024}, but shares the same two-level (middleware + POSIX OS) architecture, so our ROS~2 observations transfer directly to AP.

Linux/ROS~2 is the primary CORTS development platform, and prior work has documented its structural differences from the theoretical DAG model.
These differences are expected, since ROS~2 is a component middleware, not a DAG scheduling platform. Rather than cataloging them, we identify which platform properties invalidate DAG scheduling theory's assumptions: exactly those that system software must restore for the theory to transfer (\cref{sub:design_principles} derives the corresponding design principles).
We characterize these properties along two axes, \emph{task model} and \emph{scheduling model}, contrasting the DAG and ROS~2 sides of each.
\vspace{-0mm}
\subsection{Differences in Task Models}
\vspace{-0mm}
\subsubsection{DAG Task Model}
\vspace{-0mm}
A DAG task is represented as $G = (V, E)$, where $V$ denotes the set of vertices and $E$ denotes the set of directed edges.
Each vertex $u, v \in V$ represents a \emph{subtask}, and each edge $(u,v) \in E$ encodes a precedence constraint.
DAG tasks are released periodically or sporadically and generate an infinite sequence of jobs.
For the $k$-th job, let $a_v^k$ and $f_v^k$ denote the arrival time and the finishing time of subtask $v$, respectively.
Let the source vertex of the DAG be denoted as $v_{src}$, the precedence constraint for the $k$-th job can be formalized as
\begin{equation}\label{eq:precedence}
    \forall v \in V \setminus v_{src}:\quad a_v^k = \max_{(u,v)\in E} f_u^k.
\end{equation}
That is, $v$ becomes executable exactly when its last predecessor finishes, with no partial joins or timestamp-based merging and, ideally, zero join delay.
\vspace{-0mm}
\subsubsection{ROS~2 Task Model}
\vspace{-0mm}
A ROS~2 application operates as a chain of callbacks connected through topics.
Treating each callback as a subtask and message flow as precedence constraints, the callback chain resembles a DAG task; however, the ROS~2 API does not express these constraints directly and can inadvertently break them depending on usage.

First, ROS~2 allows \texttt{publish()} anywhere in a callback; a mid-callback publish can therefore release a successor before its predecessor has logically completed. Even when all \texttt{publish()} calls sit at the end, blocking of earlier publishes induced by quality-of-service (QoS) settings can delay later ones, again breaking precedence.
This breaks the completion boundary of~\cref{eq:precedence}, so any response-time bound that assumes edge activation at subtask completion no longer holds.
Second, multi-input callbacks require a developer-selected message-matching policy via \texttt{message\_filters}~\cite{message-filter}; common policies (\emph{ExactTime}, \emph{ApproximateTime}) join on timestamps, not the strict AND synchronization the theoretical model assumes, so which messages form one job instance, the very object DAG analyses reason about, is decided at runtime rather than statically.
Third, unless zero-copy IPC middleware (e.g., Agnocast~\cite{agnocast}) is employed, ROS~2 incurs non-negligible communication delays; the resulting join delay violates the zero-delay release assumption of~\cref{eq:precedence}, introducing an unmodeled additive term into every path-length argument.

\vspace{-0mm}
\subsection{Differences in Scheduling Models}
\vspace{-0mm}
\vspace{-0mm}
\subsubsection{DAG Scheduling Model}\label{sssec:dag_scheduling}
\vspace{-0mm}
The scheduling model centers on a ready queue of executable subtasks, into which released subtasks are immediately enqueued, and a scheduler that selects the highest-priority one using algorithms such as rate monotonic, deadline monotonic, or EDF~\cite{verucchi2023survey}.

DAG task schedulers use three preemption models~\cite{giorgio_buttazzo_b934c589, gaoyang_dai_bed08eb4}: \emph{fully preemptive} (preemption anywhere; best responsiveness, highest context-switch and cache cost), \emph{non-preemptive} (no preemption; zero overhead, blocking risk), and \emph{limited preemptive} (preemption only between subtasks; an intermediate trade-off).

Diverse dynamic algorithms have been proposed for the DAG task model (EDF-based dynamic-priority~\cite{manar_qamhieh_44b70658, sanjoy_baruah_0a3d3652}, fluid scheduling~\cite{guan-2022}, dynamic voltage and frequency scaling (DVFS)-based energy-aware~\cite{saifullah-2020, federico}, and DAG-level reservation-based scheduling~\cite{niklas-2023,holistic}), all assuming the scheduler can maintain and update DAG metadata at runtime.
\vspace{-0mm}
\subsubsection{ROS~2 Scheduling Model}\label{sssec:ros_scheduling}
\vspace{-0mm}
The executor of ROS~2 is a user-space callback scheduler nested under the OS scheduler; SingleThreadedExecutor and MultiThreadedExecutor are the defaults. Each maintains a wait set of executable instances and alternates between a \emph{polling point} and a \emph{processing window}.
At a polling point, when the wait set is empty, the lower layer is queried and one instance per activated callback (timer or subscription) is sampled; in a processing window, instances are selected non-preemptively under fixed, callback-type-determined priorities that the user cannot configure.

These defaults diverge from the DAG scheduling model~\cite{harun_teper_44b003f3}, and each divergence invalidates an assumption the theory rests on.
Instead of immediate enqueueing on release, the executor samples one instance per callback at each polling point; schedulability analyses assume a released subtask is immediately visible to the scheduler, so the polling-induced release delays add a term that none of the surveyed analyses model.
Priority comparisons occur only within the wait set, so high-priority callbacks outside it can be blocked by low-priority ones.
ROS~2 also exposes no deadline or callback-priority API, leaving DAG-level scheduling parameters no place to be expressed; thread-level OS policies exist, but nested scheduling and the absence of one-to-one callback-thread correspondence prevent precise per-instance control~\cite{CIE}.
Nested scheduling moreover means that the entity making priority decisions (the executor) is not the entity allocating CPUs (the OS scheduler), so the core premise of the DAG scheduling model, a single scheduler dispatching the globally highest-priority ready subtasks, has no counterpart on the platform.

\subsubsection{Recent Work on the ROS~2 Scheduling Model}\label{sssec:recent_work}
Recent work has removed several of these differences individually, as~\cref{tab:scheduling_model_comparison} summarizes: the EventExecutor line and RTeX eliminate polling-induced release latency through event queues with immediate enqueueing~\cite{harun_teper_44b003f3, songran_liu_641693d9}; PiCAS and its multi-threaded and dynamic-priority extensions add callback-priority interfaces~\cite{picas, RE14, RE10}; and CallbackIsolatedExecutor~\cite{CIE}, the preemptive EDF executor of Wilson et al.~\cite{physics_informed_mixed_criticality_scheduling_for_f1tenth_cars_with_preemptable_ros_2_executors_31f450b3}, and ROS\textsuperscript{RT}~\cite{ros-rt} remove nested scheduling by dedicating OS threads to callbacks.
However, all remain user-space mechanisms above a thread-based OS scheduler: each recovers individual scheduling properties, but none maintains the runtime DAG metadata that the dynamic algorithms of~\cref{sssec:dag_scheduling} presuppose, so DAG-based dynamic scheduling remains unimplementable even atop their combination.

\begin{table}[t]
    \centering
    \scriptsize
    \begin{threeparttable}
        \caption{Comparison of the ROS~2 scheduling model and the DAG scheduling model.}
        \label{tab:scheduling_model_comparison}
        \begin{tabularx}{\linewidth}{|X|c|c|c|c|c|c|}
            \hline
                                                                                                                                                    & \textbf{NLat} & \textbf{CPS} & \textbf{NNS} & \textbf{PCS} & \textbf{SRC} & \textbf{DBDS} \\ \hline

            SingleThreadedExecutor, MultiThreadedExecutor                                                                                           & ~             & ~            & ~            & ~            & ~            & ~             \\ \hline
            EventExecutor                                                                                                                           & \checkmark    & ~            & ~            & ~            & \checkmark   & ~             \\ \hline
            PiCAS~\cite{picas}                                                                                                                      & ~             & \checkmark   & ~            & ~            & ~            & ~             \\ \hline
            PiCAS Multi~\cite{RE14}                                                                                                                 & \checkmark    & \checkmark   & ~            & ~            & ~            & ~             \\ \hline
            Dynamic-Priority based Executor~\cite{RE10}                                                                                             & \checkmark    & \checkmark   & ~            & ~            & ~            & ~             \\ \hline
            Extended EventExecutor~\cite{harun_teper_44b003f3}                                                                                      & \checkmark    & \checkmark   & ~            & ~            & \checkmark   & ~             \\ \hline
            RTeX~\cite{songran_liu_641693d9}                                                                                                        & \checkmark    & \checkmark   & ~            & ~            & \checkmark   & ~             \\ \hline
            Preemptive EDF Executor~\cite{physics_informed_mixed_criticality_scheduling_for_f1tenth_cars_with_preemptable_ros_2_executors_31f450b3} & \checkmark    & \checkmark   & \checkmark   & \checkmark   & ~            & ~             \\ \hline
            CallbackIsolatedExecutor~\cite{CIE}                                                                                                     & \checkmark    & \checkmark   & \checkmark   & \checkmark   & ~            & ~             \\ \hline
            ROS\textsuperscript{RT}~\cite{ros-rt}                                                                                                   & \checkmark    & \checkmark   & \checkmark   & \checkmark   & \checkmark   & ~             \\ \hline
            DAG scheduling model                                                                                                                    & \checkmark    & \checkmark   & \checkmark   & \checkmark   & \checkmark   & \checkmark    \\ \hline
        \end{tabularx}
        \scriptsize
        \noindent
        \begin{tabular}{@{}l@{\hspace{1em}}l@{}}
            \textbf{NLat}: negligible subtask release-to-enqueue latency. & \textbf{PCS}: preemptive callback scheduling.  \\
            \textbf{CPS}: interface for callback priority setting.        & \textbf{SRC}: support for reentrant callbacks. \\
            \textbf{NNS}: non-nested scheduling.                          & \textbf{DBDS}: DAG-based dynamic scheduling.
        \end{tabular}
    \end{threeparttable}
    \vspace{-0mm}
\end{table}

\vspace{-0mm}
\section{Design Principle}\label{sub:design_principles}
\vspace{-0mm}

Building on~\cref{sec:gap}, we enumerate seven design principles (P1 to P7) for a DAG scheduling implementation platform from three perspectives.

From the \textbf{real-time scheduling researcher's perspective}:
\begin{enumerate}[label=\textbf{P\arabic*}]
    \item The platform must provide sufficient freedom to describe arbitrary DAG scheduling algorithms and their parameters.
    \item An algorithm expressed at the model level should be directly translatable into a working implementation without non-trivial platform-specific adaptations.
\end{enumerate}

From the \textbf{application programmer's perspective}:
\begin{enumerate}[resume, label=\textbf{P\arabic*}]
    \item The API must allow natural description of CORTS, maintaining continuity with the publish/subscribe paradigm widely adopted in the target ecosystem.
    \item The API must preserve the DAG task structure and all precedence constraints regardless of programming practices, so programmers can write components without scheduling expertise while the platform enforces the task model.
\end{enumerate}

From the \textbf{system perspective}:
\begin{enumerate}[resume, label=\textbf{P\arabic*}]
    \item Overheads causing divergence between model and runtime behavior (such as inter-subtask communication latency and release-to-enqueueing delay) must be minimized.
    \item The system must support every mechanism arbitrary DAG scheduling algorithms require: arbitrary preemption granularity (fully, limited, and non-preemptive), reentrant subtasks, and OS-level runtime DAG maintenance.
    \item The system must not introduce platform-induced complexity absent from the scheduling model, such as nested scheduling hierarchies that are artifacts of the system architecture.
\end{enumerate}

\vspace{-0mm}
\section{System Model}\label{sec:sys}
\vspace{-0mm}

\begin{figure}[t]
    \centerline{\includegraphics[width=\linewidth]{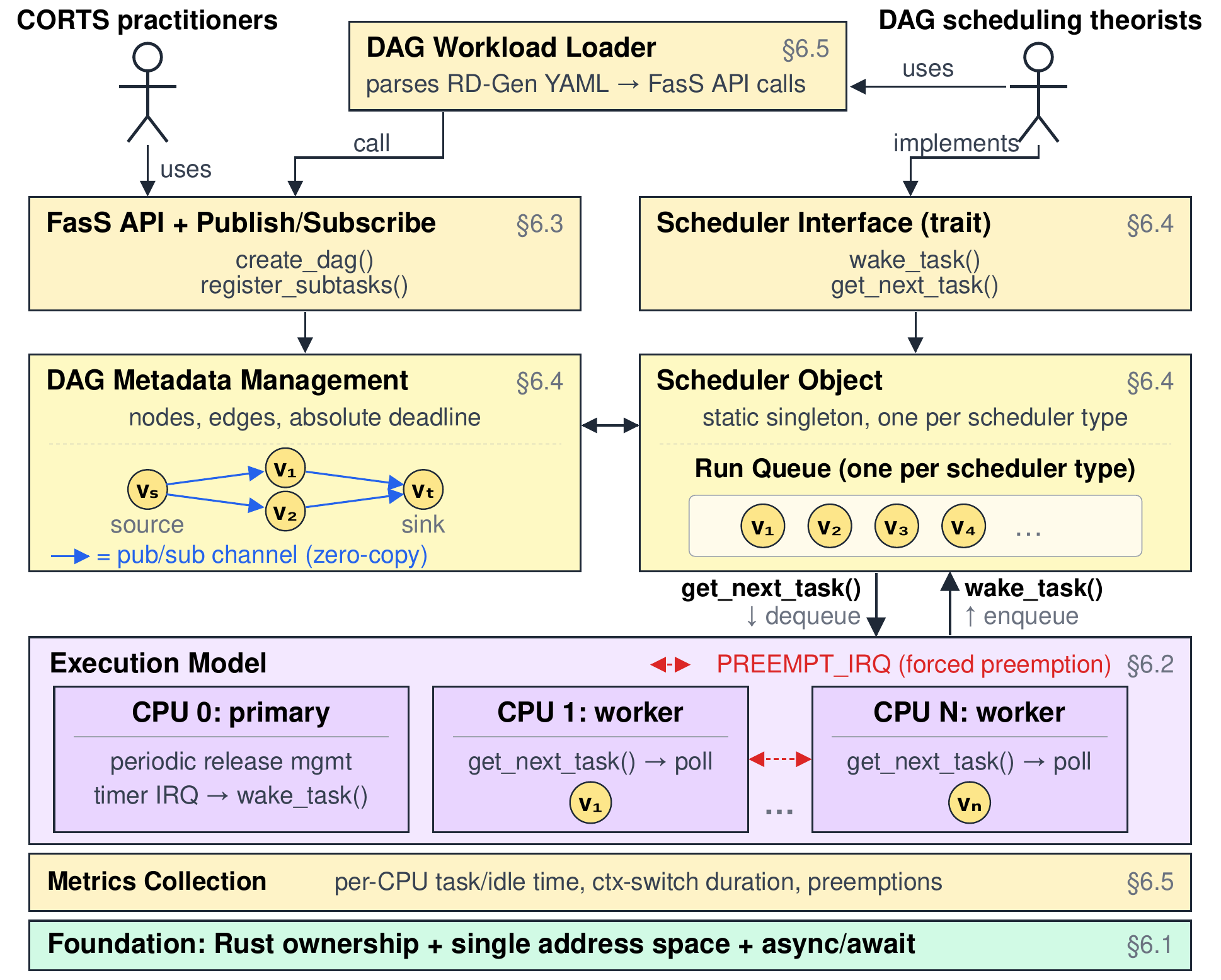}}
    \caption{Awkernel architecture. The gray tag in each block gives the subsection of~\cref{sec:design} that describes that component.}
    \label{fig:awkernel-architecture}
    \vspace{-0mm}
\end{figure}

\textbf{Overview.}
By integrating publish/subscribe and a DAG-aware scheduler into the kernel, Awkernel avoids the middleware and nested scheduling of ROS~2 (\cref{sec:gap}) while satisfying the design principles of~\cref{sub:design_principles}.
\Cref{fig:awkernel-architecture} doubles as a roadmap for~\cref{sec:design}.

\textbf{Modules.}
Awkernel comprises five components, organized by the design principles each satisfies.
(1) The \emph{foundation} layer combines the static memory safety of Rust, a single address space, and the \texttt{async/await} runtime, underpinning low-overhead inter-subtask communication and memory-safe state sharing.
(2) The \emph{execution model} comprises per-CPU workers, inter-processor interrupt (IPI)-based preemption, and periodic-task release driven by the primary CPU.
(3) The \emph{programming model} exposes the function-as-subtask (FasS) API~\cite{FasS} together with an OS-level publish/subscribe substrate.
(4) The \emph{scheduler interface} comprises the scheduler trait and the runtime DAG metadata managed inside the kernel.
(5) The \emph{evaluation support} layer provides a workload loader for YAML from RD-Gen~\cite{rd-gen} and in-kernel metrics collection exposing per-CPU latency statistics.

\textbf{Design principle coverage.}
\cref{tab:principle_mapping} maps the principles of~\cref{sub:design_principles} to the subsections of~\cref{sec:design} that satisfy them.

\begin{table}[t]
    \centering
    \caption{Mapping between design principles and Awkernel subsections.}
    \label{tab:principle_mapping}
    \scriptsize
    \setlength{\tabcolsep}{4pt}
    \begin{tabularx}{\linewidth}{|X|>{\raggedright\arraybackslash}p{0.28\linewidth}|}
        \hline
        \textbf{Design principle}                                                                & \textbf{Where satisfied}                         \\ \hline
        \textbf{P1} Freedom to describe arbitrary DAG scheduling algorithms                      & \cref{ssub:scheduler_iface}                      \\ \hline
        \textbf{P2} Model-level algorithms directly translatable                                 & \cref{ssub:scheduler_iface}                      \\ \hline
        \textbf{P3} Natural description of CORTS                                                 & \cref{ssub:programming_model}                    \\ \hline
        \textbf{P4} DAG task structure preserved regardless of programmer practice               & \cref{ssub:foundation,ssub:programming_model}    \\ \hline
        \textbf{P5} Minimization of model-runtime divergence overhead                            & \cref{ssub:foundation,ssub:execution_model}      \\ \hline
        \textbf{P6} All mechanisms for arbitrary DAG scheduling (preemption, reentrancy, DAG md) & \cref{ssub:execution_model,ssub:scheduler_iface} \\ \hline
        \textbf{P7} Absence of platform-induced complexity (no nested scheduling)                & \cref{ssub:execution_model}                      \\ \hline
    \end{tabularx}
    \vspace{-0mm}
\end{table}

\begin{table}[t]
    \centering
    \caption{Whether each design-alternative class satisfies principles P4 to P7 (\checkmark: satisfied; Limited: only with unsafe mechanisms or rewrite-scale changes; --: structurally unsatisfiable).}
    \label{tab:design_alternatives}
    \scriptsize
    \setlength{\tabcolsep}{4pt}
    \begin{tabularx}{\linewidth}{|>{\raggedright\arraybackslash}X|c|c|c|c|}
        \hline
        \textbf{Approach}                                                              & \textbf{P4}  & \textbf{P5}  & \textbf{P6}  & \textbf{P7} \\ \hline
        ROS~2 executor extensions (user space, \cref{tab:scheduling_model_comparison}) & --          & Limited     & --          & --          \\ \hline
        Linux scheduler extensions (LITMUS${ }^{\text{RT}}$, \texttt{sched\_ext})      & --          & Limited     & --          & \checkmark  \\ \hline
        Retrofitting an existing RTOS (e.g., FreeRTOS, Zephyr)                         & Limited     & Limited     & Limited     & \checkmark  \\ \hline
        Awkernel (new kernel design)                                                   & \checkmark  & \checkmark  & \checkmark  & \checkmark  \\ \hline
    \end{tabularx}
    \vspace{-0mm}
\end{table}

\textbf{Design alternatives.}
One may ask whether the design principles of~\cref{sub:design_principles} could be met by extending an existing platform instead of building a new RTOS. We argue they cannot, as~\cref{tab:design_alternatives} summarizes.
(i) User-space executor extensions of ROS~2, including all recent work summarized in~\cref{tab:scheduling_model_comparison}, remain nested under the POSIX scheduler, violating P7, and cannot make OS-level preemption decisions from DAG metadata (P6).
(ii) Extending Linux, in the style of LITMUS${ }^{\text{RT}}$~\cite{litmus-rt} or \texttt{sched\_ext}~\cite{sched-ext}, retains a thread-based scheduler abstraction with no task-graph state; more fundamentally, a POSIX kernel cannot restrict how applications publish or synchronize, so DAG semantics cannot be enforced without programmer discipline (P4), and process isolation makes zero-copy communication (P5) conditional on unsafe shared-memory mechanisms.
(iii) Retrofitting an existing RTOS would require replacing its task model, scheduler core, and communication substrate simultaneously, which amounts to a rewrite, while the memory-safe single address space that P5 relies on requires language-level ownership guarantees unavailable in C-based kernels.
A new, small, analyzable kernel is therefore not a stylistic choice but the direct consequence of P4 to P7.
Awkernel accordingly takes two positions: for theorists, it is a real-hardware testbed, playing for DAG scheduling the role that LITMUS${ }^{\text{RT}}$ has played for multiprocessor scheduling of independent tasks; for practitioners, it is not a complete replacement for production ROS~2 systems, whose ecosystem it does not replicate, but a CORTS platform maximizing timing analyzability and reuse of the DAG scheduling literature, in which following a single API contract suffices to inherit the guarantees of DAG scheduling theory.

\textbf{Use-cases for the two target users.}
Awkernel is designed for two distinct classes of users with disjoint API surfaces.
CORTS practitioners interact exclusively through the FasS API (\cref{ssub:programming_model}); they write each subtask as a pure function and never call \texttt{publish()} or manipulate scheduler state.
The kernel automatically enforces DAG semantics (completion boundaries and AND-join synchronization), so applications can safely benefit from real-time DAG scheduling without specialized knowledge of the underlying algorithm (RQ~3).
DAG scheduling theorists, in contrast, plug a new DAG scheduling algorithm into Awkernel by implementing the scheduler trait (\cref{ssub:scheduler_iface}); they reuse the FasS API and the DAG metadata facility, so the engineering effort is concentrated on the algorithm itself rather than on building an experimental platform (RQ~2).

\vspace{-0mm}
\section{Design and Implementation}\label{sec:design}
\vspace{-0mm}

\vspace{-0mm}
\subsection{Foundation: Rust, Single Address Space, and async/await}\label{ssub:foundation}
\vspace{-0mm}

The minimization-of-overhead principle (P5) demands that inter-subtask communication and the latency from release to enqueueing remain close to the model.
A single address space directly satisfies this demand: pointer-based message passing eliminates data copies, address-space switches, and IPC marshaling.

The ownership and borrow system of Rust resolves this tension by providing compile-time memory safety without any runtime mechanism, enabling the single address space to be simultaneously low-overhead and memory-safe; the same strategy has been validated by RedLeaf~\cite{redleaf} and Theseus~\cite{theseus}.

\begin{figure}[t]
    \centering
    \includegraphics[width=0.9\linewidth]{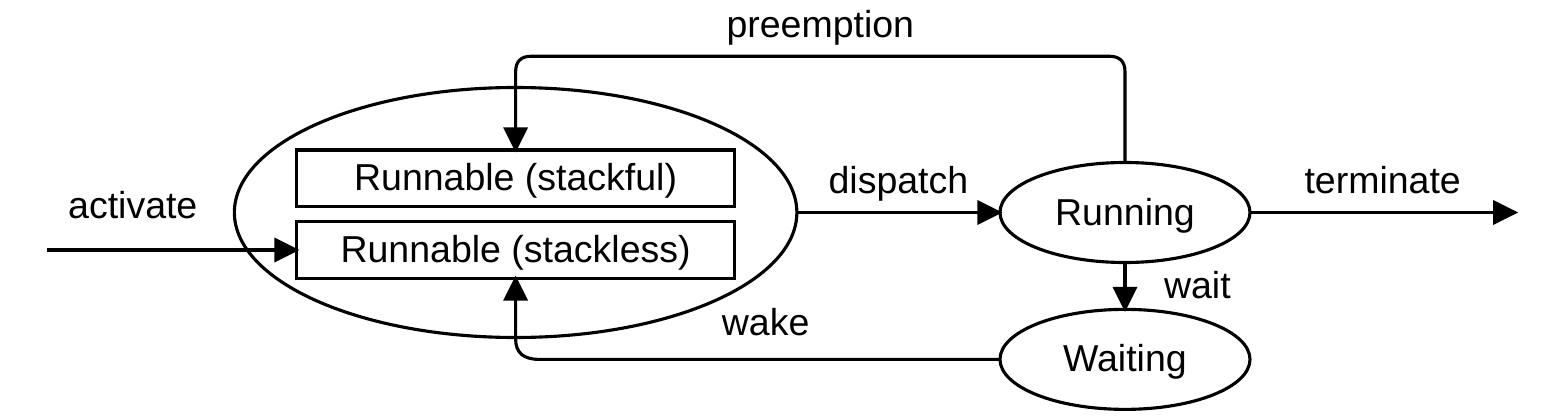}
    \vspace{-2mm}
    \caption{Task state transition.}
    \label{fig:task_state}
\end{figure}

Finally, the \texttt{async/await} runtime splits a subtask's runnable state into two flavors (\cref{fig:task_state}): a \emph{stackless} state, reached when a cooperative \texttt{.await} returns \texttt{Poll::Pending} and resumed by re-entering \texttt{poll()} with no register restoration; and a \emph{stackful} state, reached only on forced \emph{preemption} (\cref{ssub:execution_model}) and resumed by a register-restoring context switch.

\vspace{-0mm}
\subsection{Execution Model: Per-Core Workers, Preemption, and Periodic Releases}\label{ssub:execution_model}
\vspace{-0mm}

Awkernel instantiates one scheduler object per scheduling algorithm (e.g., GEDF, Prioritized FIFO, and round-robin (RR)) as a static kernel-internal singleton that owns a run queue and implements a scheduling rule; it is not a thread, has no dedicated stack, no schedulable entity, and no \texttt{spawn} site, and the precedence among scheduler objects is fixed at compile time.
Scheduler invocations occur at two opportunities: (i) on wake (by a publisher, sleep-timer expiration, or IPI), the wake site calls \texttt{wake\_task()} on the appropriate scheduler object, which enqueues the woken subtask onto the run queue of that scheduler; (ii) at the top of the run loop of each per-core worker, it calls \texttt{get\_next\_task()} to dequeue the highest-priority runnable subtask, consulting scheduler objects in the fixed precedence order.

Awkernel runs one worker per CPU, and the primary CPU (CPU~0) drives periodic releases of tasks: each iteration walks a time-ordered delta list to release expired periodic releases.
Every other CPU runs a tight \texttt{get\_next\_task() $\to$ poll} loop; thus there are $\mathit{num\_cpu}-1$ such workers.

Preemption in Awkernel is built on a dedicated IPI on \texttt{PREEMPT\_IRQ}, reserved at boot, and is exposed at the three granularities of P6.
When \texttt{wake\_task()} finds a woken subtask outranking every running and pending-preemption target across all cores, it enqueues the subtask on the preemption-pending queue of the target core and sends the IPI; the handler performs a stackful context switch to the preempting subtask.
Two reduction modes complement this default fully preemptive behavior: the \texttt{no\_preempt} Cargo feature compiles the preemption subsystem out (a non-preemptive build where \texttt{.await} yields are the only switch points), and wrapping a critical region in an \texttt{InterruptGuard}, which disables interrupts, makes it limited preemptive on demand.

A single kernel-wide MCS lock~\cite{mcs} serializes every \texttt{wake\_task()} and \texttt{get\_next\_task()} across all scheduler objects, performing in one critical section (i) the cross-core preemption decision, (ii) the run-queue enqueue/dequeue, and (iii) the per-CPU running-task table update.
This eliminates priority inversion across scheduler classes.
Subtask-level parallelism is preserved across multi-core: every subtask of every DAG task is an independent \texttt{async} task (a \texttt{Future} in Rust terminology), and each is enqueued and dispatched on its own merit; this independence is exactly what realizes reentrant subtask execution required by P6, with no need for a dedicated stack pool or worker-thread cache.

\vspace{-0mm}
\subsection{Programming Model: FasS API and Publish/Subscribe}\label{ssub:programming_model}
\vspace{-0mm}

The FasS API is a programming model that represents each subtask in the DAG task model as a simple function~$f$.
Intuitively, the arguments and return values of $f$ correspond, respectively, to the incoming and outgoing edges of the subtask in the DAG.
Strictly speaking, however, they correspond to topics (publish/subscribe data channels), so a single argument may aggregate data flowing in from multiple incoming edges, and a single return value may feed multiple outgoing edges.
The publish/subscribe (pub/sub) substrate underlying the FasS API treats each topic, within the single address space (\cref{ssub:foundation}), as a shared-memory channel that passes payloads by pointer; zero-copy is achieved by wrapping each payload in \texttt{Arc<T>}~\cite{rust-arc}.
Concretely, the API is realized as a registration interface through which the application programmer supplies the following three pieces of information for each subtask:

\begin{enumerate}
    \item \textbf{A function~$f$ representing the subtask}: within the body of $f$, operations that modify or expose the DAG structure (most notably explicit \texttt{publish()} calls) are disallowed.
    \item \textbf{Lists of topics}: the input topic list, paired with the arguments of $f$, and the output topic list, paired with its return values.
    \item \textbf{Attributes}: additional parameters required by the chosen DAG scheduling algorithm; for example, under a static-priority DAG scheduler, each FasS subtask is assigned its priority through this slot.
\end{enumerate}

The kernel wraps every registered $f$ in a control loop of the form \texttt{loop \{\,args = recv\_all.await;} \texttt{results = f(args);} \texttt{pubs.send\_all(results).await\,\}}, yielding two consequences that the programmer cannot violate.
First, outgoing publishes are structurally pinned to the tail of the subtask, matching the completion-boundary clause of the DAG task model (P4).
Second, \texttt{recv\_all.await} blocks until every input topic has supplied a message, enforcing AND-join synchronization at the API layer rather than via a developer-chosen matching policy.

The FasS API provides three registration forms that map onto the source, internal, and sink roles of a DAG task; all three carry the \texttt{attributes} slot of item~3.
The \emph{periodic} form takes a function with no input topics, an output topic list, and a period; the kernel attaches a \texttt{tick}-based reactor that releases the subtask every period.
The \emph{intermediate} form takes a function with both input and output topic lists and covers every non-source, non-sink vertex.
The \emph{sink} form takes a function whose argument tuple corresponds to its own input topics, together with a relative deadline against which the end-to-end timing of the DAG is checked.
A single call to \texttt{finish\_create\_dags(\&[dag])} then finalizes the DAG: it statically verifies the assembled graph (weakly connected, acyclic, one source, one sink) and commits it to the runtime metadata store (\cref{ssub:scheduler_iface}).
The committed DAG is immutable, which is what lets the scheduler maintain accurate per-job runtime metadata such as absolute deadlines.

\cref{lst:fass} illustrates the resulting brevity on a minimal DAG: a periodic source \texttt{f0} emitting an \texttt{i32} every \texttt{period}, an intermediate \texttt{f1} mapping it to an \texttt{(i32, i32)} pair, and a sink \texttt{f2} consuming the pair within \texttt{relative\_deadline}.
This brevity is the concrete realization of P3 (natural description) and P4 (the kernel, not the programmer, owns DAG semantics).

Awkernel does not aim to run unmodified ROS~2 binaries, and this is a consequence of the design rather than a missing feature: any API that accepts arbitrary mid-callback publishes reintroduces exactly the semantic violations identified in~\cref{sec:gap}. What Awkernel preserves is the programming model, not the API surface: components remain callbacks connected by named topics, so a port is a re-expression, not a redesign.

\begin{figure}[t]
    \begin{codebox}[Rust]
        \begin{lstlisting}[]
let dag = create_dag();

dag.register_periodic_subtask::<_, (i32,)>(
    f0, // || -> (i32,)
    vec!["topic0"], period, attributes);

dag.register_subtask::<_, (i32,), (i32, i32)>(
    f1, // |(i32,)| -> (i32, i32)
    vec!["topic0"],
    vec!["topic1", "topic2"], attributes);

dag.register_sink_subtask::<_, (i32, i32)>(
    f2, // |(i32, i32)| -> ()
    vec!["topic1", "topic2"], attributes);

finish_create_dags(&[dag.clone()]);
\end{lstlisting}
    \end{codebox}
    \vspace{-2mm}
    \caption{A minimal DAG built with the FasS API.}
    \label{lst:fass}
\end{figure}

\vspace{-0mm}
\subsection{Scheduler Interface and DAG Metadata Management}\label{ssub:scheduler_iface}
\vspace{-0mm}

Awkernel exposes the scheduling algorithm as a Rust trait that every scheduler implementation realizes.
A scheduler implementation provides \texttt{wake\_task()} and \texttt{get\_next\_task()} as required by the execution model (\cref{ssub:execution_model}), together with a priority queue keyed on the priority that the algorithm dictates.
Adding a new DAG scheduling algorithm reduces to (i) implementing the trait against an algorithm-specific run queue and (ii) registering it as a new scheduler object; the rest of the kernel (the per-CPU workers, the preemption IPI mechanism, the FasS-driven release path, and the DAG metadata store) is reused without modification.
This factoring substantiates P1 (freedom to describe arbitrary DAG scheduling algorithms) and P2 (model-level algorithms directly translatable into a working implementation).

The runtime DAG metadata maintained by the kernel mirrors the elements that a DAG scheduling algorithm needs at scheduling time.
For each DAG, the metadata records a graph identifier, the nodes (each carrying a subtask identifier and FasS-declared parameters), the edges (each tagged with its topic name), and the absolute deadline.
For each subtask, the metadata maintains a bidirectionally referenceable link to the DAG it belongs to, so a scheduler invoked on a single subtask can in $O(1)$ recover the deadline of the parent DAG, its sibling subtasks, and its sources and sinks.
The metadata is mutable at runtime through the scheduler trait (for example, GEDF rewrites the absolute deadline of a subtask at every release), so reservation-based or deadline-rewriting algorithms are first-class.
This combination of static structure (committed at \texttt{finish\_create\_dags}) and mutable per-instance state (updated at scheduling time) embodies the DAG-metadata clause of P6.

\vspace{-0mm}
\subsection{Evaluation Support: DAG Workload Loader and Metrics Collection}\label{ssub:eval_support}
\vspace{-0mm}

RD-Gen~\cite{rd-gen} is an offline random DAG generator (not part of Awkernel) that emits YAML.
Awkernel adds an in-kernel \emph{DAG workload loader} that ingests this YAML and invokes the FasS \texttt{register\_subtask} APIs, turning the synthetic workload into a runnable DAG with no manual translation.
Together with the scheduler trait (\cref{ssub:scheduler_iface}) and the FasS API (\cref{ssub:programming_model}), the loader lets a researcher evaluate a new DAG scheduling algorithm on real hardware with engineering effort proportional to the algorithm, not the platform.

Awkernel also includes a built-in metrics feature that records, on every context switch and preemption, the execution time of the preceding task, the cumulative idle time of the CPU, the context-switch duration, and the preemption count of each subtask.
Statistics are aggregated per CPU and per scheduler type and can be sampled at evaluation time.

\vspace{-0mm}
\subsection{Extensions to More General Models}\label{ssub:non_dag_and_extensions}
\vspace{-0mm}

This paper targets the single-rate DAG with AND join (\cref{sec:gap}), the model most commonly assumed in ROS~2-oriented timing analysis~\cite{casiniros, harun_teper_44b003f3, picas, RE14, RE10}; more general models (multi-rate DAGs, cause-effect chains, and time-triggered/event-driven mixtures with precedence) are widely studied. Two facts make extending this scope straightforward.

First, FasS is not the only work surface on Awkernel: application code can spawn ordinary (non-DAG) \texttt{async} tasks via \texttt{task::spawn(name, future, SchedulerType)} without attaching DAG metadata. Both task kinds share the scheduler interface of~\cref{ssub:scheduler_iface}, so any installed scheduler (Partitioned EDF, GEDF, Prioritized FIFO, or Prioritized RR) can be selected for non-DAG tasks. The \texttt{wake\_task} of GEDF, for example, derives the absolute deadline from the relative deadline carried by \texttt{SchedulerType::GEDF} whenever DAG metadata is absent, so non-DAG tasks are scheduled under deadline-driven priority with no special-case logic.

Second, inter-task communication is not bound to the FasS API. The pub/sub substrate that FasS uses is itself a public API: any task may directly instantiate publishers and subscribers to allocate ROS~2-style many-to-many topics with configurable queue depth, flow control, latching, and lifespan, independently of the DAG edges. Awkernel further exposes both synchronous mutexes (MCS/spinlock) and asynchronous mutexes (\texttt{Future}-returning sleeplock); wrapped in a reference-counted shared pointer, they provide a memory-safe shared-variable channel under the compile-time aliasing rules of Rust. Both layers offer DAG-edge-independent communication.

Combining these facts yields a sketch for workload models not natively covered. Each independent rate or event source is registered as a separate single-rate DAG, and inter-rate communication, including time-based extrapolation and partial-update compensation for stale or partial inputs, is expressed either as reads/writes of a shared variable protected by \texttt{AsyncLock<T>}, or as a directly created pub/sub topic outside any DAG edge, with the consumer DAG implementing compensation in its subtask body. First-class FasS support for these extended models, so the kernel enforces their semantics, is left to future work.

\begin{table}[t]
    \caption{Changes to DAG generation parameters in RD-Gen.}
    \label{tab:dag_parameters}
    \centering
    \scriptsize
    \setlength{\tabcolsep}{3pt}
    \begin{tabularx}{\linewidth}{|X|l||X|l|}
        \hline
        \multicolumn{2}{|c||}{\textbf{Graph structure}} & \multicolumn{2}{c|}{\textbf{Properties}}                                               \\
        \hline
        Parameter                                       & Value                                    & Parameter                 & Value           \\
        \hline
        Number of chains                                & 1 (Fixed)                                & Maximum utilization       & 1.0 (Fixed)     \\
        Number of sub-sequences                         & [1, 2, 3]                                & End-to-end deadline ratio & (1.1, 1.5, 0.1) \\
        Number of sink nodes                            & 1 (Fixed)                                & Period                    & [50, 100]       \\
        \hline
    \end{tabularx}
    \vspace{-0mm}
\end{table}

\section{Evaluation}\label{sec:evaluation}

To answer RQ1, we experimentally validate the gap analysis of~\cref{sec:gap} and the design principles of~\cref{sub:design_principles}.
Microbenchmarks verify that Awkernel satisfies the DAG scheduling model requirements of~\cref{sub:design_principles}, after which we quantify the design-space cost of expressing DAG-level dynamic scheduling on a representative POSIX-based CORTS platform (ROS~2/Linux) and on a DAG-native user-space middleware (YASMIN), compared with OS-native DAG scheduling on Awkernel.
Where~\cref{sec:gap} identified the platform properties that invalidate the assumptions of DAG scheduling theory, this section shows how those properties manifest quantitatively.
Awkernel and the entire codebase used in this evaluation are publicly available~\cite{Awkernel}.

\subsection{Evaluation Setup}\label{sub:env}
We evaluate on a Raspberry Pi~4 Model B 4\,GB. CPU~0 handles periodic release management (\cref{sec:design}), leaving three worker cores and a maximum schedulable utilization of $3.0$.
The CPU frequency is fixed at the maximum in all experiments, on every system under evaluation.
DAGs are generated with the chain-based method of RD-Gen~\cite{rd-gen} using the parameters of~\cref{tab:dag_parameters}, where $[\cdot]$ denotes a value drawn uniformly at random from the listed candidates and $(\mathit{min}, \mathit{max}, \mathit{step})$ one drawn from the arithmetic sequence it defines.
Each generated DAG consists of a single main chain (\emph{Number of chains}), from which \emph{Number of sub-sequences} branches diverge and later merge into a single sink node (\emph{Number of sink nodes}).
All time parameters are integers interpreted in milliseconds: each DAG is assigned a period drawn from \emph{Period} and a utilization capped by \emph{Maximum utilization}, and their product, the total worst-case execution time (WCET) of the DAG, is randomly partitioned into per-subtask WCETs of at least 1\,ms.
The end-to-end deadline of each DAG is its critical-path length multiplied by a ratio drawn from \emph{End-to-end deadline ratio}.
The data-size sweep in~\cref{sub:result} uses a FasS-built Awkernel DAG embedding the target size in the message; other evaluations use~\cref{tab:dag_parameters}.

\subsection{GEDF on Awkernel vs.\ POSIX-Based Realizations on Linux/ROS~2 and YASMIN}\label{sub:ros2_limits}

This section quantitatively compares three ways of realizing DAG-level dynamic scheduling for the same CORTS application: (a) Awkernel, where the OS itself maintains the DAG structure and natively schedules subtasks (P6 of~\cref{sub:design_principles}), (b) Linux/ROS~2, where DAG-level GEDF is approximated through general-purpose Linux EDF facilities (\texttt{SCHED\_DEADLINE}) combined with recent ROS~2 executor research, and (c) YASMIN, where a DAG-native middleware schedules subtasks in user space above the POSIX scheduler.
As a representative DAG-based dynamic scheduling algorithm, we consider global earliest deadline first (GEDF) for DAG tasks.

\begin{figure}[t]
    \centering
    \includegraphics[width=0.85\linewidth]{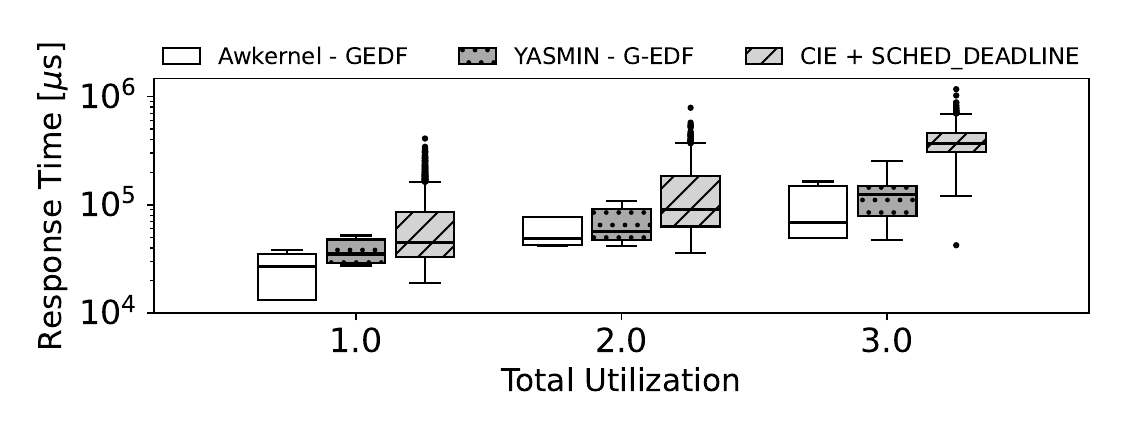}
    \caption{Response times of OS-native GEDF on Awkernel, user-space GEDF on repaired YASMIN, and the \texttt{SCHED\_DEADLINE}-based approximation of GEDF on Linux/ROS~2.}
    \label{fig-response-time}
    \vspace{-\baselineskip}
\end{figure}

\textbf{GEDF for DAG tasks.}
We briefly restate the GEDF scheduler for DAG tasks following Li et al.~\cite{GEDF}.
We consider a task set $\tau = \{\tau_1, \tau_2, \dots, \tau_n\}$ of $n$ independent sporadic tasks, where each task $\tau_i$ is represented as a DAG with period $P_i$ and relative deadline $D_i$.
Each subtask of task $\tau_i$ has its own WCET $C_i^j$.
Directed edges represent precedence constraints among subtasks within a task.

We denote by $J_{k,a}$ the $a$-th job instance of task $\tau_k$ during system execution.
Let $r_{k,a}$ and $d_{k,a}$ be the absolute release time and absolute deadline of job $J_{k,a}$, respectively, so that $d_{k,a} - r_{k,a} = D_k$.
When scheduled by a GEDF scheduler, individual subtasks carry neither release offsets nor deadlines of their own; they share the absolute deadline of the job to which they belong.
At any time $t$, the GEDF scheduler assigns all ready subtasks to available cores in non-decreasing order of absolute deadline $d_{k,a}$.

\textbf{ROS~2 workload construction.}
To evaluate whether GEDF-like scheduling can be approximated in ROS~2, we rebuild the RD-Gen workload of Awkernel~(\cref{sub:env}) as a ROS~2 application: source subtasks become \emph{timer callbacks}, triggered at the DAG period; non-source subtasks become \emph{subscription callbacks}, triggered by incoming messages.
When a subtask has multiple incoming edges, the corresponding subscription callbacks are synchronized using \texttt{message\_filters::Synchronizer} with the \emph{ExactTime} policy~\cite{message-filter}.
All callbacks are hosted within a single process to avoid inter-process communication latency and the spawning of additional DDS threads.

\textbf{Candidate scheduling approaches.}
The \texttt{SCHED\_DEADLINE} policy of Linux~\cite{sched_deadline} implements the constant bandwidth server (CBS) algorithm combined with EDF, making it a natural choice for approximating EDF-based scheduling in ROS~2.
Based on the analysis of the scheduling-model gap in~\cref{sec:gap}, we examine three approaches for scheduling the above ROS~2 system:

\begin{enumerate}
    \item \textbf{SingleThreadedExecutor with \texttt{SCHED\_DEADLINE}.}
          All callbacks forming the DAG are assigned to a single \emph{SingleThreadedExecutor}, and its thread is configured with \texttt{SCHED\_DEADLINE} using period~$= P_i$, runtime~$= C_i$ (total WCET of the DAG), and deadline~$= D_i$.

    \item \textbf{ROS\textsuperscript{RT} with \texttt{SCHED\_EXT} EDF scheduler.}
          ROS\textsuperscript{RT}~\cite{ros-rt} proposes a \texttt{SCHED\_EXT}-based EDF scheduler that applies OS-level EDF scheduling to individual ROS~2 callback threads.

    \item \textbf{CallbackIsolatedExecutor with per-thread \texttt{SCHED\_DEADLINE}.}
          Each callback's Callback Group is isolated and assigned to a \emph{CallbackIsolatedExecutor}~\cite{CIE}, which maps each callback to a dedicated thread. Each thread is then configured with \texttt{SCHED\_DEADLINE}, setting period~$= P_i$, runtime~$= C_i^j$ (the individual subtask's WCET), and deadline~$= D_i$.
\end{enumerate}

Approach~1 uses a single thread; consequently, all subtasks within a DAG execute sequentially, precluding any intra-DAG parallelism that GEDF would exploit.
Approach~2 is conceptually promising but cannot be evaluated in practice: the implementation of the proposed \texttt{SCHED\_EXT} EDF scheduler has not been publicly released, and its usage remains undocumented.
We therefore adopt Approach~3 as the ROS~2 baseline; the implementation details of GEDF on Awkernel are presented in~\cref{sec:cs_gedf}.
We stress that even Approach~3 is not an equivalent implementation of DAG-level GEDF: \texttt{SCHED\_DEADLINE} implements CBS scheduling at the thread level, and no principled mapping from DAG-level timing constraints to per-callback CBS parameters exists in the literature.
This arm of the experiment is therefore designed as evidence of a semantic mismatch and its configuration burden: it quantifies what happens when DAG-level dynamic scheduling is approximated through the best available combination of general-purpose OS facilities and ROS~2 executor research.

\textbf{YASMIN baseline.}
Beyond approximating GEDF with \texttt{SCHED\_DEADLINE}, we evaluate YASMIN~\cite{yasmin}, the middleware closest to DAG-native scheduling (\cref{subsec:rw_cort}): it maintains DAG precedence through FIFO channels and implements GEDF in user space above the POSIX scheduler.
We build the identical workload on YASMIN: source subtasks become periodic YASMIN tasks, edges become FIFO channels, and response time is measured as in the ROS~2 workload, propagating the source's start time along edges to the sink.
However, the publicly released YASMIN implementation~\cite{yasmin-impl} does not realize EDF scheduling: under its EDF configuration, no code path ever assigns a task priority (the EDF comparator is dead code), the release thread sleeps for twice the requested duration, and a semaphore leak strands ready tasks in the queue until the next release.
We repaired these defects with a patch of fewer than 30~lines and made each data-driven subtask inherit its job's absolute deadline, so that the repaired YASMIN schedules with exactly the GEDF semantics defined above; all modifications strengthen the baseline.
Mirroring the Awkernel setup, three \texttt{SCHED\_FIFO} worker threads run on cores~1--3 and the release thread on core~0.
This arm of the experiment thus complements Approach~3: it preserves the GEDF semantics and instead isolates the architectural cost of executing them in user space above a DAG-unaware POSIX scheduler.

\textbf{Response-time comparison.}
\Cref{fig-response-time} compares the response times of GEDF on Awkernel, Approach~3 (CallbackIsolatedExecutor with per-thread \texttt{SCHED\_DEADLINE}), and the repaired YASMIN across different total utilization levels.
As the figure shows, the response times of Approach~3 inflate substantially in a large number of cases.
This inflation is not a deficiency of \texttt{SCHED\_DEADLINE} itself but the manifestation of the semantic mismatch stated above.
First, CBS assumes that each task is released sporadically with a known minimum inter-arrival time, whereas subscription callbacks are activated by message arrivals whose spacing is perturbed by scheduling delays, queuing effects, and precedence-induced waiting; when a callback is activated earlier than the CBS mechanism expects, the server defers its deadline replenishment, inflating the observed response time.
Second, the per-callback CBS parameters have no principled derivation from the DAG-level deadline, so any chosen configuration, including ours, misrepresents the DAG-level constraint to some degree.
Awkernel does not exhibit this inflation because its OS scheduler operates directly on the DAG-level deadline (P6 of~\cref{sub:design_principles}).

The repaired YASMIN, which executes the same GEDF logic in user space, comes considerably closer to Awkernel, keeping its response-time tails bounded up to a total utilization of $2.5$.
Nevertheless, the median response times of YASMIN remain $11$--$81\%$ above those of Awkernel, and the gap widens toward saturation: at a total utilization of $3.0$, YASMIN yields $1.8\times$ the median response time.
The residual cost is architectural: even with identical scheduling logic, YASMIN dispatches subtasks through a user-space ready queue, realizes preemption with POSIX signals, and mediates every release through a periodic scheduler thread, all above a POSIX scheduler that remains unaware of DAG precedence.
The comparison thus characterizes two distinct costs of expressing DAG-level GEDF above a POSIX scheduler, relative to a design in which the OS itself maintains the DAG structure and schedules subtasks accordingly: the semantic mismatch and configuration burden of the thread-level CBS approximation, and the architectural overhead of user-space dispatching even when the GEDF semantics are preserved.
In particular, the Linux/ROS~2 results should be read as the price of an approximation, not as a ranking between two equivalent GEDF implementations.

\subsection{Awkernel Microbenchmark Results}\label{sub:result}
From~\cref{sub:design_principles}, we quantitatively evaluate the two measurable principles, \emph{negligible communication latency between subtasks} and \emph{negligible release-to-enqueue latency} (P5), plus two scheduler metrics: preemption and context-switch overhead.
All use the workload of~\cref{sub:ros2_limits}: three concurrent RD-Gen DAGs (\cref{tab:dag_parameters}) with total utilization swept from $1.0$ to $3.0$, spanning lightly loaded to saturated operation of the three worker cores.

\begin{figure}[t]
    \begin{minipage}[t]{0.48\linewidth}
        \centering
        \includegraphics[width=\linewidth]{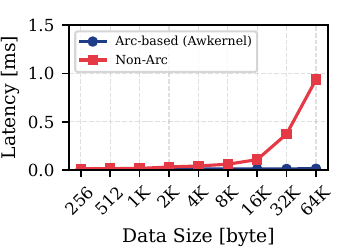}
        \caption{Medians of the publish/subscribe communication latency between subtasks by data size.}
        \label{fig:pubsub}
    \end{minipage}\hfill
    \begin{minipage}[t]{0.48\linewidth}
        \centering
        \includegraphics[width=\linewidth]{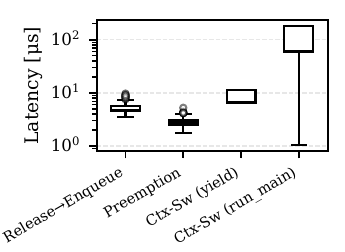}
        \caption{Latency breakdown of basic scheduling operations.}
        \label{fig:basic-latency}
    \end{minipage}
\end{figure}

\textbf{Communication latency between subtasks.}
\Cref{fig:pubsub} compares publish/subscribe latency using Arc (pointer-based sharing) versus raw copying. With Arc, the single address space of Awkernel enables zero-copy and keeps latency constant in data size; without Arc, cloning makes latency grow linearly.

\textbf{Scheduling-path overheads.}
\Cref{fig:basic-latency} breaks down three latencies on the scheduling path. (i) \emph{Release-to-enqueue}: time from a subtask becoming executable to its insertion into the run queue, stays in the microsecond range, with fluctuations from lock-acquisition waits during DAG metadata access, so Awkernel satisfies P5 of~\cref{sub:design_principles}. (ii) \emph{Preemption decision}: also in the microsecond range, adding negligible delay to the scheduling path. (iii) \emph{Context switch}: Awkernel has two paths: \emph{yield} only saves/restores registers, whereas \emph{run\_main} additionally checks for a saved context (restoring it if present), yielding slightly higher latency.

For large real-world payloads,~\cref{sec:cs_mrm} confirms end-to-end impact remains negligible and comparable to existing systems.

\textbf{Bounded blocking under load.}
A single kernel-wide MCS lock serializes all scheduler operations (wake-up, enqueue, dequeue, and the per-CPU \emph{RUNNING}-table update) within a fixed four-step critical section, bounding worker-side blocking by the number of contending workers, independent of DAG or subtask count. Sweeping utilization from $1.0$ to $3.0$ keeps median release-to-enqueue latency in the microsecond range (\cref{fig:basic-latency}) with no blocking explosion at saturation; scaling beyond four cores is out of scope, as we target the Raspberry Pi~4 embedded class.

\section{Case Study}\label{sec:casestudy}
Two case studies validate Awkernel: \cref{sec:cs_gedf} implements GEDF, showing theorists can evaluate a DAG algorithm on real hardware with minimal effort (RQ2, Contribution~2); \cref{sec:cs_mrm} partially ports an Autoware application, showing practitioners can leverage DAG scheduling via publish/subscribe and the FasS API (RQ3, Contribution~3).

\subsection{Case Study 1: Implementation of GEDF}\label{sec:cs_gedf}
Using the customizable scheduler of~\cref{sec:design}, adding an algorithm requires registering a unique name, declaring priority queues, and inserting algorithm-specific logic into the task-insertion routine of the scheduler trait; other operations (e.g., preemption checks, task removal) are reused with minor edits.

On enqueue (\cref{alg1}), GEDF records $r_{i,k}$, computes $d_{i,k}=r_{i,k}+D_i$ using $D_i$ from the DAG sink node, and after a mutex-protected priority update either preempts via an interrupt or queues the task. The core logic is \textbf{226}~lines in a single file.

\begin{algorithm}[t]
    \caption{GEDF core logic.}
    \label{alg1}
    \begin{algorithmic}[1]
        \STATE wake$\_$time $\leftarrow$ CURRENT$\_$TIME;
        \STATE absolute$\_$deadline $\leftarrow$ wake$\_$time $+$ relative$\_$deadline (retrieved from DAG sink node);
        \STATE Update priority based on absolute$\_$deadline;
    \end{algorithmic}
\end{algorithm}

\begin{figure}[t]
    \centering
    \includegraphics[width=\linewidth, trim=18 14 18 30, clip, keepaspectratio]{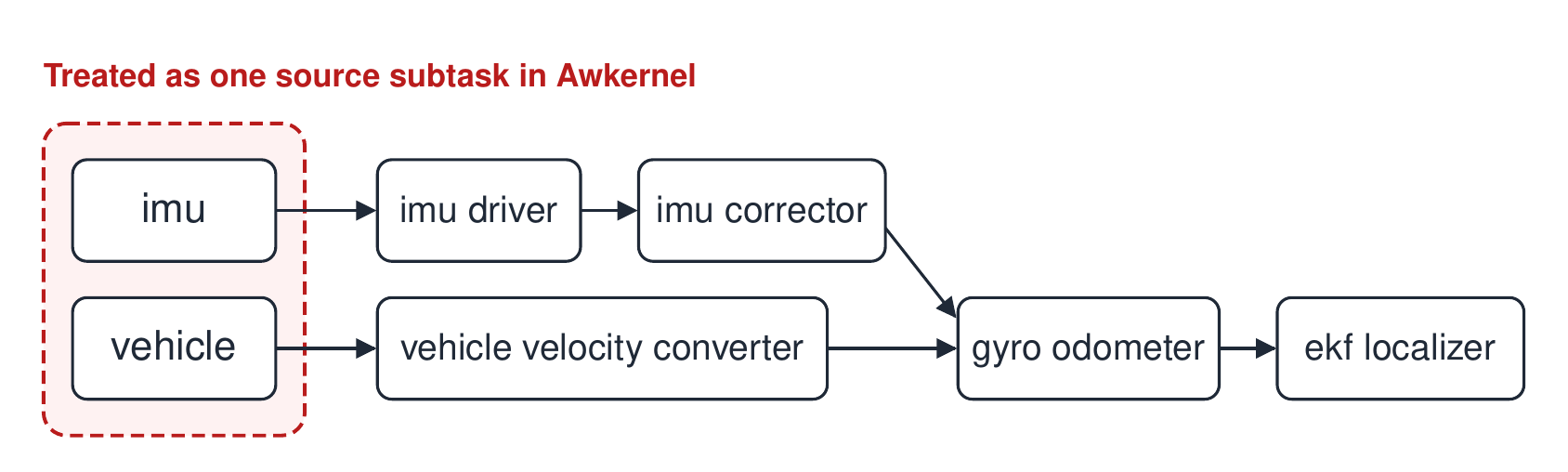}
    \caption{Minimum risk maneuver (MRM) in Autoware.}
    \label{fig:mrm-autoware}
\end{figure}

\begin{figure}[t]
    \centering
    \includegraphics[width=\linewidth, keepaspectratio]{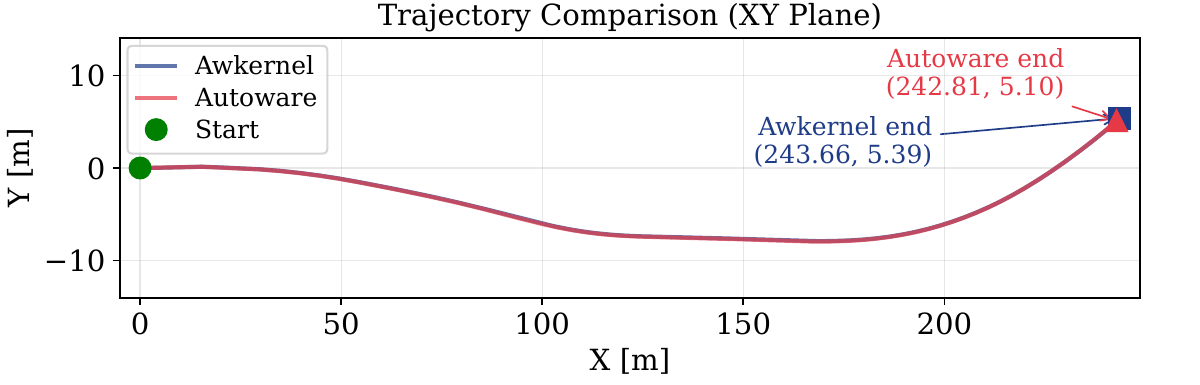}
    \caption{Trajectory comparison between Autoware and Awkernel.}
    \label{fig:mrm-tra}
\end{figure}

\subsection{Case Study 2: Partial Porting of the Autoware Application}\label{sec:cs_mrm}
Case study~2 ports part of Autoware~\cite{autoware}, a representative ROS~2/Linux CORTS application, onto Awkernel via the FasS API to show that FasS-enforced DAG semantics preserve operational behavior.
The scope is \emph{portability}: the two systems ingest sensor data through different I/O paths (CSV replay over UDP/IP versus the live AWSIM stack), so a timing comparison would confound platform and I/O effects; the controlled timing comparison on identical workloads is in~\cref{sec:evaluation}.

MRM, an Autoware feature that safely stops the vehicle when humans cannot intervene, forms a minimal DAG suited to our platform. In~\cref{fig:mrm-autoware}, \texttt{imu driver} and \texttt{vehicle velocity converter} forward angular velocity and velocity downstream each cycle~\cite{multi_dead}, and \texttt{ekf localizer} self-localizes from the odometry.

\subsubsection{Implementation Details}
Each MRM node is reimplemented as a subtask of one DAG via the FasS API~\cite{FasS}, following a fixed mechanical mapping: each ROS~2 node becomes one FasS function, each subscription an argument, each publish a return value, and each timer a periodic release. Inputs are CSVs from AWSIM ros$\_$bag data; the self-position of \texttt{ekf localizer} is sent over UDP/IP and qualitatively compared with an Autoware reference trace.

\vspace{1mm}

\subsubsection{Evaluation of MRM}
\Cref{fig:mrm-tra} shows the ported MRM tracks the Autoware reference within \textbf{0.29\,m} in $y$, well inside the MRM safety envelope. The residual is by design: Autoware activates \texttt{gyro odometer} and \texttt{ekf localizer} via messages or timers, while Awkernel fires a subtask only when all predecessors deliver data under FasS-enforced AND synchronization.
Autoware-style timer-based join or partial update can be expressed on Awkernel (\cref{ssub:non_dag_and_extensions}). We avoid this here to keep the case study within the strict single-rate DAG with AND-join scope.

\vspace{-0mm}
\section{Conclusion}\label{sec:conclusion}
\vspace{-0mm}
This paper introduced Awkernel, the first open-source RTOS with native real-time DAG scheduling support: a real-hardware testbed for theorists, and for practitioners a CORTS platform maximizing timing analyzability and reuse of the DAG scheduling literature rather than a replacement for ROS~2.
We identified gaps between the DAG model and POSIX-based CORTS, derived seven design principles, and validated them on Raspberry Pi~4: microsecond-range overheads, zero-copy Arc communication, and FasS-enforced precedence by construction (\textbf{RQ1}). Baselines on Linux/ROS~2 (\texttt{SCHED\_DEADLINE}-based approximation) and YASMIN (repaired user-space GEDF) exposed, respectively, the mismatch and configuration burden of approximating DAG-level GEDF with thread-level CBS, and the cost of user-space execution.
The GEDF case study implemented a DAG algorithm in 226~lines via the scheduler interface, showing low implementation effort for theorists (\textbf{RQ2}). The Autoware MRM case study ported a real-world CORTS application via the FasS API, tracking the Autoware reference within 0.29\,m under FasS-enforced semantics (\textbf{RQ3}).
These insights, API-enforced DAG semantics and OS-level DAG metadata management, can also be fed back into ROS~2 and AP.

\textbf{Future work.}
First, we will extend the FasS API beyond single-rate DAGs with AND-join to multi-rate DAGs, cause-effect chains, and mixed time-triggered/event-driven nodes with precedence (\cref{ssub:non_dag_and_extensions}). Second, we will implement and compare additional DAG scheduling algorithms (federated~\cite{li-2014}, reservation-based~\cite{exact}, and fluid~\cite{guan-2022}) on Awkernel for reproducible real-hardware comparisons. Third, we plan a tool that generates FasS subtask skeletons from ROS~2 node graphs.

\vspace{1mm}

\ifanonymous\else
    \begin{credits}
        \subsubsection{\ackname}
        This work was supported by JST CREST Grant Number \mbox{JPMJCR23M1} and based
        on results obtained from a project, Green Innovation Fund Projects /
        Development of In-vehicle Computing and Simulation Technology for
        Energy Saving in Electric Vehicles (JPNP21027), subsidized by the New
        Energy and Industrial Technology Development Organization (NEDO).

        \subsubsection{\discintname}
        A. Yano, Y. Kobayashi, Y. Takano, and T. Ishikawa-Aso are employed by
        TIER IV, Inc., which develops Awkernel as open-source software. The
        other authors have no competing interests to declare that are relevant
        to the content of this article.
    \end{credits}
\fi

\bibliographystyle{splncs04unsrt}
\bibliography{reference}

@misc{Awkernel,
    key = {Awkernel},
    title="{Awkernel} and the codebase used in this evaluation",
    howpublished = "Repository withheld for double-blind review; it will be released in the camera-ready version"
}

@misc{message-filter,
    key = {message-filter},
    title="{message\_filters} --- {ROS} {Wiki}",
    howpublished="\url{https://wiki.ros.org/message_filters}"
}

@misc{sched-ext,
    key = {sched-ext},
    title="Extensible Scheduler Class --- The {Linux} Kernel Documentation",
    howpublished="\url{https://docs.kernel.org/scheduler/sched-ext.html}"
}

@misc{rust-arc,
    key = {Rust-Arc},
    title="Struct {alloc::sync::Arc}",
    howpublished =
    "\url{https://doc.rust-lang.org/beta/alloc/sync/struct.Arc.html}"
}

@misc{autoware,
    key = {Autoware},
    title="{Autoware}",
    howpublished="\url{https://github.com/autowarefoundation/autoware}"}

@ARTICLE{RE10,
author={Al Arafat, Abdullah and Wilson, Kurt and Yang, Kecheng and Guo, Zhishan},
journal={IEEE Transactions on Computer-Aided Design of Integrated Circuits and Systems},
title={Dynamic Priority Scheduling of Multithreaded {ROS} 2 Executor With Shared Resources},
year={2024},
volume={43},
number={11},
pages={3732-3743},
doi={10.1109/TCAD.2024.3445259},
}

@inproceedings{RE14,
title={Timing Analysis and Priority-driven Enhancements of {ROS} 2 Multi-threaded Executors},
author={Hoora Sobhani and Hyunjong Choi and Hyoseung Kim},
booktitle={Proceedings of the 29th IEEE Real-Time and Embedded Technology and Applications Symposium (RTAS)},
year={2023},
pages={106-118},
doi = {10.1109/RTAS58335.2023.00016},
}

@INPROCEEDINGS{CIE,
  author={Ishikawa-Aso, Takahiro and Yano, Atsushi and Azumi, Takuya and Kato, Shinpei},
  booktitle={Proceedings of the 31st IEEE Real-Time and Embedded Technology and Applications Symposium (RTAS)}, 
  title={Work in Progress: Middleware-Transparent Callback Enforcement in Commoditized Component-Oriented Real-Time Systems}, 
  year={2025},
  pages={426-429},
  doi={10.1109/RTAS65571.2025.00017},
}

@INPROCEEDINGS{FasS,
  author={Ishikawa-Aso, Takahiro and Yano, Atsushi and Kobayashi, Yutaro and Jin, Takumi and Takano, Yuuki and Kato, Shinpei},
  booktitle={Proceedings of the 46th IEEE Real-Time Systems Symposium (RTSS)}, 
  title={Work-in-Progress: Function-as-Subtask {API} Replacing Publish/Subscribe for {OS}-Native {DAG} Scheduling}, 
  year={2025},
  pages={604-607},
  doi = {10.1109/RTSS66672.2025.00059},
}

@misc{ros2_repo,
  key = {ROS2},
  title = "{Open Source Robotics Foundation (OSRF)}: {ROS}~2",
  howpublished = "\url{https://github.com/ros2}"
}

@INPROCEEDINGS{ros2_paper,
  author={Maruyama, Yuya and Kato, Shinpei and Azumi, Takuya},
  booktitle={Proceedings of the 25th International Conference on Embedded Software (EMSOFT)}, 
  title={Exploring the performance of {ROS2}},
  year={2016},
  volume={},
  number={},
  pages={5:1--5:10},
  doi={10.1145/2968478.2968502},
}

@article{verucchi2023survey,
  title={A survey on real-time {DAG} scheduling, revisiting the global-partitioned infinity war},
  author={Verucchi, Micaela and Olmedo, Ignacio Sa{\~n}udo and Bertogna, Marko},
  journal={Real-Time Systems},
  volume={59},
  number={3},
  pages={479--530},
  year={2023},
  publisher={Springer},
  doi = {10.1007/s11241-023-09403-3}
}

@INPROCEEDINGS{GEDF,
  author={Li, Jing and Agrawal, Kunal and Lu, Chenyang and Gill, Christopher},
  booktitle={Proceedings of the 25th Euromicro Conference on Real-Time Systems (ECRTS)}, 
  title={Analysis of Global {EDF} for Parallel Tasks}, 
  year={2013},
  pages={3-13},
  doi={10.1109/ECRTS.2013.12},
}

@ARTICLE{adaptive1,
  author={Senapati, Debabrata and Sarkar, Arnab and Karfa, Chandan},
  journal={IEEE Transactions on Computer-Aided Design of Integrated Circuits and Systems}, 
  title={Energy-Aware Real-Time Scheduling of Multiple Periodic {DAGs} on Heterogeneous Systems}, 
  year={2023},
  volume={42},
  number={8},
  pages={2447-2460},
  doi={10.1109/TCAD.2022.3228504},
}

@inproceedings{casiniros,
 author = {Casini, Daniel and Bla{\ss}, Tobias and L{\"u}tkebohle, Ingo and Brandenburg, Bj{\"o}rn},
 year = {2019},
 booktitle={Proceedings of the 31st Euromicro Conference on Real-Time Systems (ECRTS)},
 title = {Response-Time Analysis of {ROS} 2 Processing Chains Under Reservation-Based Scheduling},
 pages={6:1--6:23},
 doi = {10.4230/LIPIcs.ECRTS.2019.6}
}

@INPROCEEDINGS{ros_multi,
  author={Jiang, Xu and Ji, Dong and Guan, Nan and Li, Ruoxiang and Tang, Yue and Yi, Wang},
  booktitle={Proceedings of the 43rd IEEE Real-Time Systems Symposium (RTSS)}, 
  title={Real-Time Scheduling and Analysis of Processing Chains on Multi-threaded Executor in {ROS} 2}, 
  year={2022},
  volume={},
  number={},
  pages={27-39},
  doi={10.1109/RTSS55097.2022.00013},
}

@INPROCEEDINGS{ros_stv,
  author={Bla{\ss}, Tobias and Casini, Daniel and Bozhko, Sergey and Brandenburg, Bj{\"o}rn B.},
  booktitle={Proceedings of the 42nd IEEE Real-Time Systems Symposium (RTSS)}, 
  title={A {ROS} 2 Response-Time Analysis Exploiting Starvation Freedom and Execution-Time Variance}, 
  year={2021},
  volume={},
  number={},
  pages={41-53},
  doi={10.1109/RTSS52674.2021.00016},
}

@INPROCEEDINGS{ros_e2e,
  author={Teper, Harun and G{\"u}nzel, Mario and Ueter, Niklas and von der Br{\"u}ggen, Georg and Chen, Jian-Jia},
  booktitle={Proceedings of the 43rd IEEE Real-Time Systems Symposium (RTSS)}, 
  title={End-To-End Timing Analysis in {ROS2}}, 
  year={2022},
  volume={},
  number={},
  pages={53-65},
  doi={10.1109/RTSS55097.2022.00015},
}

@INPROCEEDINGS{ros_sync,
  author={Li, Ruoxiang and Guan, Nan and Jiang, Xu and Guo, Zhishan and Dong, Zheng and Lv, Mingsong},
  booktitle={Proceedings of the 43rd IEEE Real-Time Systems Symposium (RTSS)}, 
  title={Worst-Case Time Disparity Analysis of Message Synchronization in {ROS}}, 
  year={2022},
  volume={},
  number={},
  pages={40-52},
  doi={10.1109/RTSS55097.2022.00014},
}

@INPROCEEDINGS{multi_dead,
  author={Yano, Atsushi and Azumi, Takuya},
  booktitle={Proceedings of the 45th IEEE Real-Time Systems Symposium (RTSS)}, 
  title={Work-in-Progress: Multi-Deadline {DAG} Scheduling Model for Autonomous Driving Systems}, 
  year={2024},
  volume={},
  number={},
  pages={451-454},
  doi={10.1109/RTSS62706.2024.00049},
}

@INPROCEEDINGS{rd-gen,
  author={Yano, Atsushi and Azumi, Takuya},
  booktitle={Proceedings of the 26th IEEE International Symposium on Real-Time Distributed Computing (ISORC)}, 
  title={{RD-Gen}: Random {DAG} Generator Considering Multi-rate Applications for Reproducible Scheduling Evaluation}, 
  year={2023},
  pages={21-31},
  doi={10.1109/ISORC58943.2023.00015},
}

@INPROCEEDINGS{ros_single_pri,
  author={Tang, Yue and Feng, Zhiwei and Guan, Nan and Jiang, Xu and Lv, Mingsong and Deng, Qingxu and Yi, Wang},
  booktitle={Proceedings of the 41st IEEE Real-Time Systems Symposium (RTSS)}, 
  title={Response Time Analysis and Priority Assignment of Processing Chains on {ROS2} Executors}, 
  year={2020},
  volume={},
  number={},
  pages={231-243},
  doi={10.1109/RTSS49844.2020.00030},
}

@INPROCEEDINGS{litmus-rt,
  author={Calandrino, John M. and Leontyev, Hennadiy and Block, Aaron and Devi, UmaMaheswari C. and Anderson, James H.},
  booktitle={Proceedings of the 27th IEEE Real-Time Systems Symposium (RTSS)}, 
  title={{LITMUS\textsuperscript{RT}}: A Testbed for Empirically Comparing Real-Time Multiprocessor Schedulers}, 
  year={2006},
  volume={},
  number={},
  pages={111-126},
  doi={10.1109/RTSS.2006.27},
 }

@inproceedings{picas,
  title={{PiCAS}: New Design of Priority-Driven Chain-Aware Scheduling for {ROS2}},
  author={Hyunjong Choi and Yecheng Xiang and Hyoseung Kim},
  booktitle={Proceedings of the 27th IEEE Real-Time and Embedded Technology and Applications Symposium (RTAS)},
  year={2021},
  pages={251--263},
  doi={10.1109/RTAS52030.2021.00028},
}

@inproceedings{ros-rt,
  author       = {Sizhe Liu and Rohan Wagle and
                  Shareef Ahmed and Zelin Tong and
                  James Anderson},
  title        = {{ROS\textsuperscript{RT}}: Enabling Flexible Scheduling in {ROS}~2},
  booktitle    = {Proceedings of the 46th IEEE Real-Time Systems Symposium (RTSS)},
  pages        = {42--54},
  year         = {2025},
  doi = {10.1109/RTSS66672.2025.00013}
}

@INPROCEEDINGS{junjie,
  author={Shi, Junjie and G{\"u}nzel, Mario and Ueter, Niklas and von der Br{\"u}ggen, Georg and Chen, Jian-Jia},
  booktitle={Proceedings of the 30th IEEE Real-Time and Embedded Technology and Applications Symposium (RTAS)}, 
  title={{DAG} Scheduling with Execution Groups}, 
  year={2024},
  pages={149-160},
  doi={10.1109/RTAS61025.2024.00020},
}

@INPROCEEDINGS{han,
  author={Han, Jongwoo and Park, Seonghyeon and Jeon, Haejoo and Lee, Chang-Gun},
  booktitle={Proceedings of the 28th IEEE Real-Time and Embedded Technology and Applications Symposium (RTAS)}, 
  title={Guaranteeing Safety Despite Physical Errors in Cyber-Physical Systems}, 
  year={2022},
  pages={1-12},
  doi={10.1109/RTAS54340.2022.00009},
}

@INPROCEEDINGS{pass,
  author={Zhao, Hui and Li, Shangshu and Wang, Quan and Wang, Jing},
  booktitle={Proceedings of the 27th IEEE Real-Time and Embedded Technology and Applications Symposium (RTAS)}, 
  title={Work in Progress: Power-aware Scheduling Strategy for Multiple {DAGs} in the Heterogeneous Cloud}, 
  year={2021},
  pages={509-512},
  doi={10.1109/RTAS52030.2021.00064},
}

@INPROCEEDINGS{micaela,
  author={Verucchi, Micaela and Theile, Mirco and Caccamo, Marco and Bertogna, Marko},
  booktitle={Proceedings of the 26th IEEE Real-Time and Embedded Technology and Applications Symposium (RTAS)}, 
  title={Latency-Aware Generation of Single-Rate {DAGs} from Multi-Rate Task Sets}, 
  year={2020},
  pages={226-238},
  doi={10.1109/RTAS48715.2020.000-4},
}

@INPROCEEDINGS{guo,
  author={Guo, Zhishan and Bhuiyan, Ashikahmed and Liu, Di and Khan, Aamir and Saifullah, Abusayeed and Guan, Nan},
  booktitle={Proceedings of the 25th IEEE Real-Time and Embedded Technology and Applications Symposium (RTAS)}, 
  title={Energy-Efficient Real-Time Scheduling of {DAGs} on Clustered Multi-Core Platforms}, 
  year={2019},
  pages={156-168},
  doi={10.1109/RTAS.2019.00021},
}

@INPROCEEDINGS{saud,
  author={Wasly, Saud and Pellizzoni, Rodolfo},
  booktitle={Proceedings of the 25th IEEE Real-Time and Embedded Technology and Applications Symposium (RTAS)}, 
  title={Bundled Scheduling of Parallel Real-Time Tasks}, 
  year={2019},
  pages={130-142},
  doi={10.1109/RTAS.2019.00019},
}

@INPROCEEDINGS{petros,
  author={Voudouris, Petros and Stenstr{\"o}m, Per and Pathan, Risat},
  booktitle={Proceedings of the 37th IEEE Real-Time Systems Symposium (RTSS)}, 
  title={Timing-anomaly free dynamic scheduling of task-based parallel applications}, 
  year={2016},
  pages={371},
  doi={10.1109/RTSS.2016.050},
}

@inproceedings{ahmed,
    author = {Shareef Ahmed and James H. Anderson},
    title = {A Soft-Real-Time Optimal Scheduler for {DAG}
Tasks with Node-Level Self Dependencies},
    booktitle = {Proceedings of the 46th IEEE Real-Time Systems Symposium (RTSS)},
    pages = {122-134},
    year = {2025},
    doi = {10.1109/RTSS66672.2025.00019}
}

@INPROCEEDINGS{guan,
  author={Guan, Fei and Lee, Jinkyu and Xue, Chun Jason and Wu, Jen-Ming and Guan, Nan},
  booktitle={Proceedings of the 45th IEEE Real-Time Systems Symposium (RTSS)},
  title={Mixed-Criticality Federated Scheduling for Relaxed-Deadline {DAG} Tasks},
  year={2024},
  pages={362-374},
  doi={10.1109/RTSS62706.2024.00038},
}

@ARTICLE{guan-2022,
  author={Guan, Fei and Peng, Long and Qiao, Jiaqing},
  journal={IEEE Transactions on Computers},
  title={A Fluid Scheduling Algorithm for {DAG} Tasks With Constrained or Arbitrary Deadlines},
  year={2022},
  volume={71},
  number={8},
  pages={1860-1873},
  doi={10.1109/TC.2021.3111512},
}

@INPROCEEDINGS{holistic,
  author={Tong, Zelin and Ahmed, Shareef and Anderson, James H.},
  booktitle={Proceedings of the 44th IEEE Real-Time Systems Symposium (RTSS)}, 
  title={Holistically Budgeting Processing Graphs}, 
  year={2023},
  pages={27-39},
  doi={10.1109/RTSS59052.2023.00013}
}

@ARTICLE{ching-chi,
  author={Lin, Ching-Chi and Shi, Junjie and Ueter, Niklas and G{\"u}nzel, Mario and Reineke, Jan and Chen, Jian-Jia},
  journal={IEEE Transactions on Computers}, 
  title={Type-Aware Federated Scheduling for Typed {DAG} Tasks on Heterogeneous Multicore Platforms}, 
  volume={72},
  number={5},
  year={2023},
  pages={1286-1300},
  doi={10.1109/TC.2022.3202748},
}

@ARTICLE{niklas-2023,
  author={Ueter, Niklas and G{\"u}nzel, Mario and von der Br{\"u}ggen, Georg and Chen, Jian-Jia},
  journal={IEEE Transactions on Computers}, 
  title={Parallel Path Progression {DAG} Scheduling}, 
  year={2023},
  volume={72},
  number={10},
  pages={3002-3016},
  doi = {10.1109/TC.2023.3280137},
}

@INPROCEEDINGS{yue-2023,
  author={Tang, Yue and Jiang, Xu and Guan, Nan and Liu, Songran and Luo, Xiantong and Yi, Wang},
  booktitle={Proceedings of the 44th IEEE Real-Time Systems Symposium (RTSS)}, 
  title={Optimizing End-to-End Latency of Sporadic Cause-Effect Chains Using Priority Inheritance}, 
  year={2023},
  volume={},
  number={},
  pages={411-422},
  doi={10.1109/RTSS59052.2023.00042},
}

@INPROCEEDINGS{exact,
  author={Ahmed, Shareef and Anderson, James H.},
  booktitle={Proceedings of the 43rd IEEE Real-Time Systems Symposium (RTSS)}, 
  title={Exact Response-Time Bounds of Periodic {DAG} Tasks under Server-Based Global Scheduling}, 
  year={2022},
  pages={447-459},
  doi={10.1109/RTSS55097.2022.00045},
}

@INPROCEEDINGS{qingqiang,
  author={He, Qingqiang and Guan, Nan and Lv, Mingsong and Jiang, Xu and Chang, Wanli},
  booktitle={Proceedings of the 43rd IEEE Real-Time Systems Symposium (RTSS)}, 
  title={Bounding the Response Time of {DAG} Tasks Using Long Paths}, 
  year={2022},
  pages={474-486},
  doi={10.1109/RTSS55097.2022.00047},
}

@INPROCEEDINGS{xu-2021,
  author={Jiang, Xu and Guan, Nan and Liang, Haochun and Tang, Yue and Qiao, Lei and Yi, Wang},
  booktitle={Proceedings of the 42nd IEEE Real-Time Systems Symposium (RTSS)}, 
  title={Virtually-Federated Scheduling of Parallel Real-Time Tasks}, 
  year={2021},
  pages={482-494},
  doi={10.1109/RTSS52674.2021.00050},
}

@INPROCEEDINGS{federico,
  author={Reghenzani, Federico and Bhuiyan, Ashikahmed and Fornaciari, William and Guo, Zhishan},
  booktitle={Proceedings of the 42nd IEEE Real-Time Systems Symposium (RTSS)}, 
  title={A Multi-Level {DPM} Approach for Real-Time {DAG} Tasks in Heterogeneous Processors}, 
  year={2021},
  pages={14-26},
  doi={10.1109/RTSS52674.2021.00014},
}

@INPROCEEDINGS{shuai,
  author={Zhao, Shuai and Dai, Xiaotian and Bate, Iain and Burns, Alan and Chang, Wanli},
  booktitle={Proceedings of the 41st IEEE Real-Time Systems Symposium (RTSS)}, 
  title={{DAG} Scheduling and Analysis on Multiprocessor Systems: Exploitation of Parallelism and Dependency}, 
  year={2020},
  pages={128-140},
  doi={10.1109/RTSS49844.2020.00022},
}

@INPROCEEDINGS{zheng,
  author={Dong, Zheng and Liu, Cong},
  booktitle={Proceedings of the 40th IEEE Real-Time Systems Symposium (RTSS)}, 
  title={An Efficient Utilization-Based Test for Scheduling Hard Real-Time Sporadic {DAG} Task Systems on Multiprocessors}, 
  year={2019},
  volume={},
  number={},
  pages={181-193},
  doi={10.1109/RTSS46320.2019.00026},
}

@INPROCEEDINGS{casini-2018,
  author={Casini, Daniel and Biondi, Alessandro and Nelissen, Geoffrey and Buttazzo, Giorgio},
  booktitle={Proceedings of the 39th IEEE Real-Time Systems Symposium (RTSS)}, 
  title={Partitioned Fixed-Priority Scheduling of Parallel Tasks Without Preemptions}, 
  year={2018},
  volume={},
  number={},
  pages={421-433},
  doi={10.1109/RTSS.2018.00056},
}

@INPROCEEDINGS{xu-2017,
  author={Jiang, Xu and Guan, Nan and Long, Xiang and Yi, Wang},
  booktitle={Proceedings of the 38th IEEE Real-Time Systems Symposium (RTSS)}, 
  title={Semi-Federated Scheduling of Parallel Real-Time Tasks on Multiprocessors}, 
  year={2017},
  volume={},
  number={},
  pages={80-91},
  doi={10.1109/RTSS.2017.00015},
}

@INPROCEEDINGS{xu-2016,
  author={Jiang, Xu and Long, Xiang and Guan, Nan and Wan, Han},
  booktitle={Proceedings of the 37th IEEE Real-Time Systems Symposium (RTSS)}, 
  title={On the Decomposition-Based Global {EDF} Scheduling of Parallel Real-Time Tasks}, 
  year={2016},
  volume={},
  number={},
  pages={237-246},
  doi={10.1109/RTSS.2016.031},
}

@INPROCEEDINGS{li-2014,
  author={Li, Jing and Chen, Jian-Jia and Agrawal, Kunal and Lu, Chenyang and Gill, Chris and Saifullah, Abusayeed},
  booktitle={Proceedings of the 26th Euromicro Conference on Real-Time Systems (ECRTS)}, 
  title={Analysis of Federated and Global Scheduling for Parallel Real-Time Tasks}, 
  year={2014},
  volume={},
  number={},
  pages={85-96},
  doi={10.1109/ECRTS.2014.23},
}

@InProceedings{kunal-2018,
  author =	{Agrawal, Kunal and Baruah, Sanjoy},
  title =	{A Measurement-Based Model for Parallel Real-Time Tasks},
  booktitle =	{Proceedings of the 30th Euromicro Conference on Real-Time Systems (ECRTS)},
  pages =	{5:1--5:19},
  year =	{2018},
  doi =	{10.4230/LIPIcs.ECRTS.2018.5},
}

@InProceedings{qingqiang-2021,
  author =	{He, Qingqiang and Lv, Mingsong and Guan, Nan},
  title =	{Response Time Bounds for {DAG} Tasks with Arbitrary Intra-Task Priority Assignment},
  booktitle =	{Proceedings of the 33rd Euromicro Conference on Real-Time Systems (ECRTS)},
  pages =	{8:1--8:21},
  year =	{2021},
  doi =	{10.4230/LIPIcs.ECRTS.2021.8},
}

@InProceedings{pathan-2018,
  author =	{Pathan, Risat Mahmud},
  title =	{Improving the Schedulability and Quality of Service for Federated Scheduling of Parallel Mixed-Criticality Tasks on Multiprocessors},
  booktitle =	{Proceedings of the 30th Euromicro Conference on Real-Time Systems (ECRTS)},
  pages =	{12:1--12:22},
  year =	{2018},
  doi =	{10.4230/LIPIcs.ECRTS.2018.12},
}

@InProceedings{saifullah-2020,
  author =	{Saifullah, Abusayeed and Fahmida, Sezana and Modekurthy, Venkata P. and Fisher, Nathan and Guo, Zhishan},
  title =	{{CPU} Energy-Aware Parallel Real-Time Scheduling},
  booktitle =	{Proceedings of the 32nd Euromicro Conference on Real-Time Systems (ECRTS)},
  pages =	{2:1--2:26},
  year =	{2020},
  doi =	{10.4230/LIPIcs.ECRTS.2020.2},
}

@InProceedings{aromolo-2023,
  author =	{Aromolo, Federico and Nelissen, Geoffrey and Biondi, Alessandro},
  title =	{Replication-Based Scheduling of Parallel Real-Time Tasks},
  booktitle =	{Proceedings of the 35th Euromicro Conference on Real-Time Systems (ECRTS)},
  pages =	{18:1--18:23},
  year =	{2023},
  doi =	{10.4230/LIPIcs.ECRTS.2023.18},
}

@InProceedings{agnocast,
  author= {Takahiro Ishikawa-Aso and Shinpei Kato},
  title= {{ROS} 2 {Agnocast}: Supporting Unsized Message Types for True Zero-Copy Publish/Subscribe {IPC}},
  booktitle= {Proceedings of the 28th IEEE International Symposium on Real-Time Distributed Computing (ISORC)},
  pages= {1-10},
  year={2025},
  doi = {10.1109/ISORC65339.2025.00026}
}

@article{giorgio_buttazzo_b934c589,
  author = {Giorgio Buttazzo and Marko Bertogna and Gang Yao},
  title = {Limited Preemptive Scheduling for Real-Time Systems. A Survey},
  journal = {IEEE Transactions on Industrial Informatics},
  volume = {9},
  number = {1},
  pages = {3-15},
  year = {2013},
  doi={10.1109/TII.2012.2188805},
}

@article{gaoyang_dai_bed08eb4,
  author = {Gaoyang Dai and Morteza Mohaqeqi and Petros Voudouris and Wang Yi},
  title = {Response-Time Analysis of Limited-Preemptive Sporadic {DAG} Tasks},
  journal = {IEEE Transactions on Computer-Aided Design of Integrated Circuits and Systems},
  volume = {41},
  number = {11},
  pages = {3673-3684},
  year = {2022},
  doi = {10.1109/TCAD.2022.3200527},
}

@inproceedings{manar_qamhieh_44b70658,
  author = {Manar Qamhieh and Fr\'{e}d\'{e}ric Fauberteau and Laurent George and
   Serge Midonnet},
  booktitle={Proceedings of the 21st ACM International Conference on Real-Time Networks and Systems (RTNS)},
  title = {Global {EDF} scheduling of directed acyclic graphs on multiprocessor systems},
  pages = {287-296},
  year = {2013},
  doi = {10.1145/2516821.2516836},
}

@inproceedings{sanjoy_baruah_0a3d3652,
  author = {Sanjoy Baruah and Vincenzo Bonifaci and Alberto Marchetti-Spaccamela},
  title = {The Global {EDF} Scheduling of Systems of Conditional Sporadic {DAG} Tasks},
  booktitle={Proceedings of the 27th Euromicro Conference on Real-Time Systems (ECRTS)},
  pages = {222-231},
  year = {2015},
  doi={10.1109/ECRTS.2015.27},
}

@inproceedings{harun_teper_44b003f3,
  author = {Teper, Harun and Bell, Oren and G\"{u}nzel, Mario and Gill, Chris and Chen, Jian-Jia},
  title = {Reconciling {ROS} 2 with Classical Real-Time Scheduling of Periodic Tasks},
  booktitle = {Proceedings of the 31st IEEE Real-Time and Embedded Technology and Applications Symposium (RTAS)},
  pages = {177-189},
  year = {2025},
  doi = {10.1109/RTAS65571.2025.00027},
}

@article{songran_liu_641693d9,
  author = {Songran Liu and Xu Jiang and Nan Guan and Zilong Wang and Minghe Yu and Wang Yi},
  title = {{RTeX}: An Efficient and Timing-Predictable Multithreaded Executor for {ROS} 2},
  journal = {IEEE Transactions on Computer-Aided Design of Integrated Circuits and Systems},
  volume = {43},
  number = {9},
  pages = {2578-2591},
  year = {2024},
  doi={10.1109/TCAD.2024.3380551}
}

@inproceedings{physics_informed_mixed_criticality_scheduling_for_f1tenth_cars_with_preemptable_ros_2_executors_31f450b3,
  author = {Wilson, Kurt and Al Arafat, Abdullah and Baugh, John and Yu, Ruozhou and Guo, Zhishan},
  title = {Physics-Informed Mixed-Criticality Scheduling for {F1Tenth} Cars with Preemptable {ROS} 2 Executors},
  booktitle = {Proceedings of the 31st IEEE Real-Time and Embedded Technology and Applications Symposium (RTAS)},
  year = {2025},
  pages = {215-227},
  doi = {10.1109/RTAS65571.2025.00030},
}

@article{sched_deadline,
  author = {Lelli, Juri and Scordino, Claudio and Abeni, Luca and Faggioli, Dario},
  title = {Deadline scheduling in the {Linux} kernel},
  journal = {Software: Practice and Experience},
  volume = {46},
  number = {6},
  pages = {821--839},
  year = {2016},
  doi = {10.1002/spe.2335},
}

@article{sun-omptg-2022,
  author  = {Sun, Jinghao and Jin, Tao and Xue, Yekai and Zhang, Liwei and Liu, Jinrong and Guan, Nan and Zhou, Quan},
  title   = {{ompTG}: From {OpenMP} Programs to Task Graphs},
  journal = {Journal of Systems Architecture},
  volume  = {126},
  pages   = {102470},
  year    = {2022},
  doi     = {10.1016/j.sysarc.2022.102470}
}

@inproceedings{sun-openmp-wcrt-2021,
  author    = {Sun, Jinghao and Guan, Nan and Guo, Zhishan and Xue, Yekai and He, Jing and Tan, Guozhen},
  title     = {Calculating Worst-Case Response Time Bounds for {OpenMP} Programs with Loop Structures},
  booktitle = {Proceedings of the 42nd IEEE Real-Time Systems Symposium (RTSS)},
  year      = {2021},
  pages     = {123--135},
  doi       = {10.1109/RTSS52674.2021.00022}
}

@article{sun-openmp-conditional-2020,
  author  = {Sun, Jinghao and Guan, Nan and Sun, Jingchang and Zhang, Xi and Chi, Yaoyao and Li, Feng},
  title   = {Algorithms for Computing the {WCRT} Bound of {OpenMP} Task Systems with Conditional Branches},
  journal = {IEEE Transactions on Computers},
  volume  = {70},
  number  = {1},
  pages   = {57--71},
  year    = {2021},
  doi     = {10.1109/TC.2020.2984502}
}

@inbook{serrano-openmp-2022,
author = {Serrano, Maria A. and Royuela, Sara and Marongiu, Andrea and Quiñones, Eduardo},
year = {2022},
month = {09},
pages = {33-62},
title = {Predictable Parallel Programming with OpenMP},
isbn = {9781003338413},
doi = {10.1201/9781003338413-3}
}

@inproceedings{fuerst-autosar-ap-2016,
  author    = {F\"urst, Simon and Bechter, Markus},
  title     = {{AUTOSAR} for Connected and Autonomous Vehicles: The {AUTOSAR} Adaptive Platform},
  booktitle = {Proceedings of the 46th IEEE/IFIP International Conference on Dependable Systems and Networks Workshop (DSN-W)},
  year      = {2016},
  pages     = {215--217},
  doi       = {10.1109/DSN-W.2016.24}
}

@inproceedings{bellassai-slelet-2025,
  author    = {Bellassai, Davide and Sciangula, Gerlando and Scordino, Claudio and Casini, Daniel and Biondi, Alessandro},
  title     = {Modeling the {SL-LET} Paradigm in {AUTOSAR} Adaptive},
  booktitle = {Proceedings of the Design, Automation \& Test in Europe Conference (DATE)},
  year      = {2025},
  pages     = {1--7},
  doi       = {10.23919/DATE64628.2025.10992699}
}

@article{bellassai-let-posix-2023,
  author  = {Bellassai, Davide and Biondi, Alessandro and Biasci, Alessandro and Morelli, Bruno},
  title   = {Supporting Logical Execution Time in Multi-Core {POSIX} Systems},
  journal = {Journal of Systems Architecture},
  volume  = {144},
  pages   = {102987},
  year    = {2023},
  doi     = {10.1016/j.sysarc.2023.102987}
}

@inproceedings{dasari-qnx-2022,
  author    = {Dasari, Dakshina and Becker, Matthias and Casini, Daniel and Bla{\ss}, Tobias},
  title     = {End-to-End Analysis of Event Chains under the {QNX} Adaptive Partitioning Scheduler},
  booktitle = {Proceedings of the 28th IEEE Real-Time and Embedded Technology and Applications Symposium (RTAS)},
  year      = {2022},
  pages     = {214--227},
  doi       = {10.1109/RTAS54340.2022.00025}
}

@inproceedings{henle-ros2-vs-ap-2022,
  author    = {Henle, Jacqueline and Stoffel, Martin and Schindewolf, Marc and N{\"a}gele, Ann-Therese and Sax, Eric},
  title     = {Architecture Platforms for Future Vehicles: A Comparison of {ROS2} and Adaptive {AUTOSAR}},
  booktitle = {Proceedings of the 25th IEEE International Conference on Intelligent Transportation Systems (ITSC)},
  year      = {2022},
  pages     = {3095--3102},
  doi       = {10.1109/ITSC55140.2022.9921894}
}

@inproceedings{iwakami-ap-ros2-2024,
  author    = {Iwakami, Ryudai and Peng, Bo and Hanyu, Hiroyuki and Ishigooka, Tasuku and Azumi, Takuya},
  title     = {{AUTOSAR AP} and {ROS 2} Collaboration Framework},
  booktitle = {Proceedings of the 27th Euromicro Conference on Digital System Design (DSD)},
  year      = {2024},
  pages     = {319--326},
  doi       = {10.1109/DSD64264.2024.00050}
}

@inproceedings{redleaf,
  title     = {{RedLeaf}: Isolation and Communication in a Safe Operating System},
  author    = {Narayanan, Vikram and Huang, Tianjiao and Detweiler, David and Appel, Dan and Li, Zhaofeng and Zellweger, Gerd and Burtsev, Anton},
  booktitle = {Proceedings of the 14th USENIX Symposium on Operating Systems Design and Implementation (OSDI)},
  pages     = {21--39},
  year      = {2020}
}

@inproceedings{theseus,
  title     = {{Theseus}: an Experiment in Operating System Structure and State Management},
  author    = {Boos, Kevin and Liyanage, Namitha and Ijaz, Ramla and Zhong, Lin},
  booktitle = {Proceedings of the 14th USENIX Symposium on Operating Systems Design and Implementation (OSDI)},
  pages     = {1--19},
  year      = {2020}
}

@article{mcs,
  title     = {Algorithms for Scalable Synchronization on Shared-Memory Multiprocessors},
  author    = {Mellor-Crummey, John M. and Scott, Michael L.},
  journal   = {ACM Transactions on Computer Systems (TOCS)},
  volume    = {9},
  number    = {1},
  pages     = {21--65},
  year      = {1991},
  doi       = {10.1145/103727.103729}
}

@inproceedings{yasmin,
  author    = {Rouxel, Benjamin and Altmeyer, Sebastian and Grelck, Clemens},
  title     = {{YASMIN}: a Real-Time Middleware for {COTS} Heterogeneous Platforms},
  booktitle = {Proceedings of the 22nd International Middleware Conference (Middleware)},
  year      = {2021},
  pages     = {298--309},
  publisher = {ACM},
  address   = {New York, NY, USA},
  doi       = {10.1145/3464298.3493402}
}

@misc{yasmin-impl,
    key = {yasmin-impl},
    title="{YASMIN} implementation",
    howpublished="\url{https://bitbucket.org/uva-sne/coordinationruntime}"
}

\end{document}